\documentclass[11pt]{article}

\usepackage[margin=1in]{geometry}
\usepackage[utf8]{inputenc}
\usepackage[T1]{fontenc}
\usepackage{amsmath, amssymb, amsthm}
\usepackage{graphicx}
\usepackage{authblk}      
\usepackage{tabularx}
\usepackage{multirow}
\usepackage{booktabs}
\usepackage{hyperref}
\usepackage{xurl}
\usepackage{xr-hyper}
\usepackage{soul}
\usepackage[normalem]{ulem}
\usepackage{xcolor}
\usepackage{tcolorbox}     
\usepackage{natbib}  
\usepackage{pdfpages}      

\graphicspath{{Fig/}}

\theoremstyle{plain}

\theoremstyle{definition}

\theoremstyle{remark}

\title{Evolution and determinants of firm-level systemic risk in local production networks}

\author[1,2,*]{Anna Mancini}
\author[3,4,5]{Balázs Lengyel}
\author[6,7]{Riccardo Di Clemente}
\author[1,2,8]{Giulio Cimini}

\affil[1]{Physics Department and INFN, University of Rome Tor Vergata, 00133 Rome, Italy}
\affil[2]{Centro Ricerche Enrico Fermi, 00184 Rome, Italy}
\affil[3]{Agglomeration, Networks and Innovation MTA Momentum Research Group, HUN-REN Centre for Economic and Regional Studies, 1093 Budapest, Hungary}
\affil[4]{ANETI Lab, Corvinus Institute for Advanced Studies, Corvinus University of Budapest, 1093 Budapest, Hungary}
\affil[5]{Department of Network Science, Institute for Data Analytics and Information Systems, Corvinus University of Budapest, 1093 Budapest, Hungary}
\affil[6]{Complex Connections Lab, Network Science Institute, Northeastern University, E1W1LP London, United Kingdom}
\affil[7]{ISI Foundation, 10126 Turin, Italy}
\affil[8]{Institute for Complex Systems, National Research Council, 00185 Rome, Italy}
\affil[*]{Corresponding author: \href{mailto:anna.mancini@cref.it}{anna.mancini@cref.it}}

\date{}   

\begin{document}
	
	\maketitle
	
	\begin{abstract}
		Recent crises like the Covid-19 pandemic and geopolitical tensions have exposed vulnerabilities and caused disruptions of supply chains, leading to product shortages, increased costs, and economic instability. This has prompted growing efforts to assess systemic risk, namely the effects of firm disruptions on entire economies.
		However, the ability of firms to react to crises by rewiring their supply links has been largely overlooked, limiting our understanding of production networks resilience.
		Here, we study dynamics and determinants of firm-level systemic risk in the Hungarian economy from 2015 to 2022. We benchmark our results to a heuristic maximum entropy null model that generates randomized production networks while preserving the total input (demand) and output (supply) of each firm at the sector level. 
		We show that the fairly stable set of firms with highest systemic risk undergoes a structural change during Covid-19, as those enabling economic exchanges become key players in the economy -- a pattern not reproduced by the null model.
		Although empirical systemic risk closely matches the null value prior to the pandemic, it becomes significantly lower afterwards, reflecting the emergence of a more resilient economy driven by firms’ adaptive behavior. Furthermore, firms' international trade volume (being itself a channel of potential disruption) becomes a significant predictor of their systemic risk. However, international linkages alone cannot fully explain the observed trends, as imports and exports exert opposing effects on local systemic risk through the supply and demand channels.
		
	\end{abstract}
	
	\begin{tcolorbox}[colback=gray!5, colframe=gray!40, title=Significance Statement]
		Production networks underpin modern economies, yet their adaptive response to major disruptions remains poorly understood. By analyzing firm-level data from the Hungarian economy over 2015-2022, we reveal how production networks dynamically reconfigured in response to the Covid-19 pandemic shock. The resulting rewiring of supply connections ultimately led to a more resilient economy, both in absolute terms and relative to configurations generated by a maximum-entropy network model with firm-level productivity constraints. Firms facilitating economic exchanges became key players of the economy, while international trade volume turned into a significant predictor of systemic risk. Our findings provide new insights into the determinants of firm-level systemic risk, with potential implications for the assessments of economic resilience at the firm and system level.
	\end{tcolorbox}

\section{Introduction}
Production networks arise as firms exchange goods and services that are needed for their own production \cite{atalay2011network,Lorincz:2024aa}. The resulting interdependencies between suppliers, manufacturers, and consumers across industries and geographic regions foster economic and technological progress and are essential to the functioning of modern economies \cite{choi2006supply,McNerney:2022aa, foreign_structural_change}. 
However, production networks have become increasingly vulnerable \cite{Ivanov:2021aa}, as a consequence of the growing globalization, interconnectedness, and complexity of supply chains \cite{choi2001supply, CHENG20142328, surana2005,Kim:2015aa,Brintrup:2017aa} as well as the constant drive toward efficiency \cite{just-in-time,kannan2005just}. 
Interconnectedness, in particular, implies that disruptions affecting firms in one region or sector can ripple across the entire network, with large-scale economic consequences \cite{barrot2016input,inoue2019firm,Aldrighetti:2021aa}.
The propagation of shocks through global supply chains has occurred repeatedly in recent years.
For example, natural disasters such as Hurricane Katrina and the Fukushima earthquake had economic impacts far beyond the area directly affected \cite{inoue2023disruption, carvalho2021jap, kashiwagi2021propagation}. Similarly, geopolitical events, such as the Russian invasion of Ukraine, disrupted global grain supplies for importing nations \cite{ukrinvasion,ukraine2025korovkin}.
In the paramount case of the Covid-19 pandemic, lockdown measures and the subsequent factory shutdowns had a disruptive effect on global production networks, affecting entire economies that struggled to supply basic goods to the population \cite{guan2020global, chowdhury2021covid, pichler2020production, pichler2022simultaneous}. These events have stimulated growing interest in assessing systemic risk -- specifically, understanding how disruptions propagate through production networks and generate large-scale economic effects, and how local networks face external shocks through their export and import connections. 

%SYSTEMIC RISK
Although the propagation of economic shocks has been traditionally studied using industry-level Input-Output tables \cite{leontief1986input, Miller:2009aa, Lee2019transmission, Klimek:2019aa, contreras2014propagation}, recent research has highlighted the role of granular network effects \cite{acemoglu2012net, carvalho2019production, Gabaix:2011aa}. As a consequence, attention has increasingly shifted towards production networks at the firm level 
\cite{fujiwara2010large,ohnishi2010network,mizuno2014structure,brintrup2015nested,kito2018disentangling,Brintrup:2018aa,inoue2019firm,peydro2020production,mattsson2021functional,Wiedmer:2021aa,diem2022quantifying,pichler2023building,chakraborty2024inequality,reisch2025rewiring, tabachova2024estimating}, also thanks to the growing availability of large-scale datasets of inter-firm relationships \cite{dyne2015belgian,borsos2020unfolding,bacilieri2022firm}. Notably, the use of firm-level data allows to overcome the aggregation bias inherent in Input-Output tables \cite{Morimoto:1970aa}, thereby enabling more reliable estimates of systemic risk \cite{diem2024estimating}. The supply chain management literature studied several aspects of disruption propagation along supply chains, such as the {\em supply chain resilience} \cite{craighead2007severity}, the {\em snowball effect} \cite{swierczek2014impact}, the {\em ripple effect} \cite{ivanov2014ripple,ivanov2017simulation,dolgui2018ripple} and the {\em nexus suppliers} index \cite{yan2015theory,shao2018data}, recognizing the importance of considering variability \cite{kaki2015disruptions} and topological features \cite{ledwoch2018systemic} of the underlying network.
Furthermore, how supplier and customer disruptions affect firms depends on both their production process and network position. The recently proposed \emph{Economic Systemic Risk Index} (ESRI) \cite{diem2022quantifying} builds on these insights to quantify the risk associated with a firm in terms of the total reduction in economic output caused by its failure.

%Adaptive response
Measuring systemic risk typically relies on a static network assumption, where firms and their relationships within the production network are fixed as shocks propagate. 
However, when a crisis occurs, firms actively respond by adapting their supply chains \cite{WANG2020214, BELHADI2021120447, KLOCKNER2023113664, zhao_CAS, Konig:2022aa}. 
These adaptations may include finding alternative suppliers, changing logistics pathways, reallocating production, or exploiting new market opportunities that arise during the crisis \cite{bode_AMJ, bode_DS,KRAMMER2022102368, sima_IJLRA}. 
Such adaptive responses are not merely temporary fixes, but can lead to long-term modifications in the structure of the production network \cite{9174793, doi:10.1177/00081256211073355}. 
These structural changes can then either mitigate systemic risk, by making the network more resilient and diversified, or exacerbate it, by creating new dependencies and bottlenecks. 
As a result, systemic risk can no longer be regarded as a static property that is solely determined by the pre-crisis network structure. 

In this work, we study the evolution and determinants of firm-level systemic risk using a subset of the Hungarian production network (see Methods), which contains all firms (above a minimum size) located in Budapest and all their value-added tax (VAT) transactions (exceeding a minimum threshold), 
during years from 2015 to 2022 -- thus including the outbreak of the Covid-19 pandemic.
We compute and track the ESRI value for each firm in the network (see Methods) to draw a detailed portrait of the evolution of systemic risk at both the individual firm and sector level over an extended time span. This enables us to investigate 
how resilience evolves over time, providing a comprehensive analysis of system-level adaptation to the Covid-19 shock and identifying new key players of the economy.

\begin{figure*}[!ht]
	\centering
	\includegraphics[width=\textwidth]{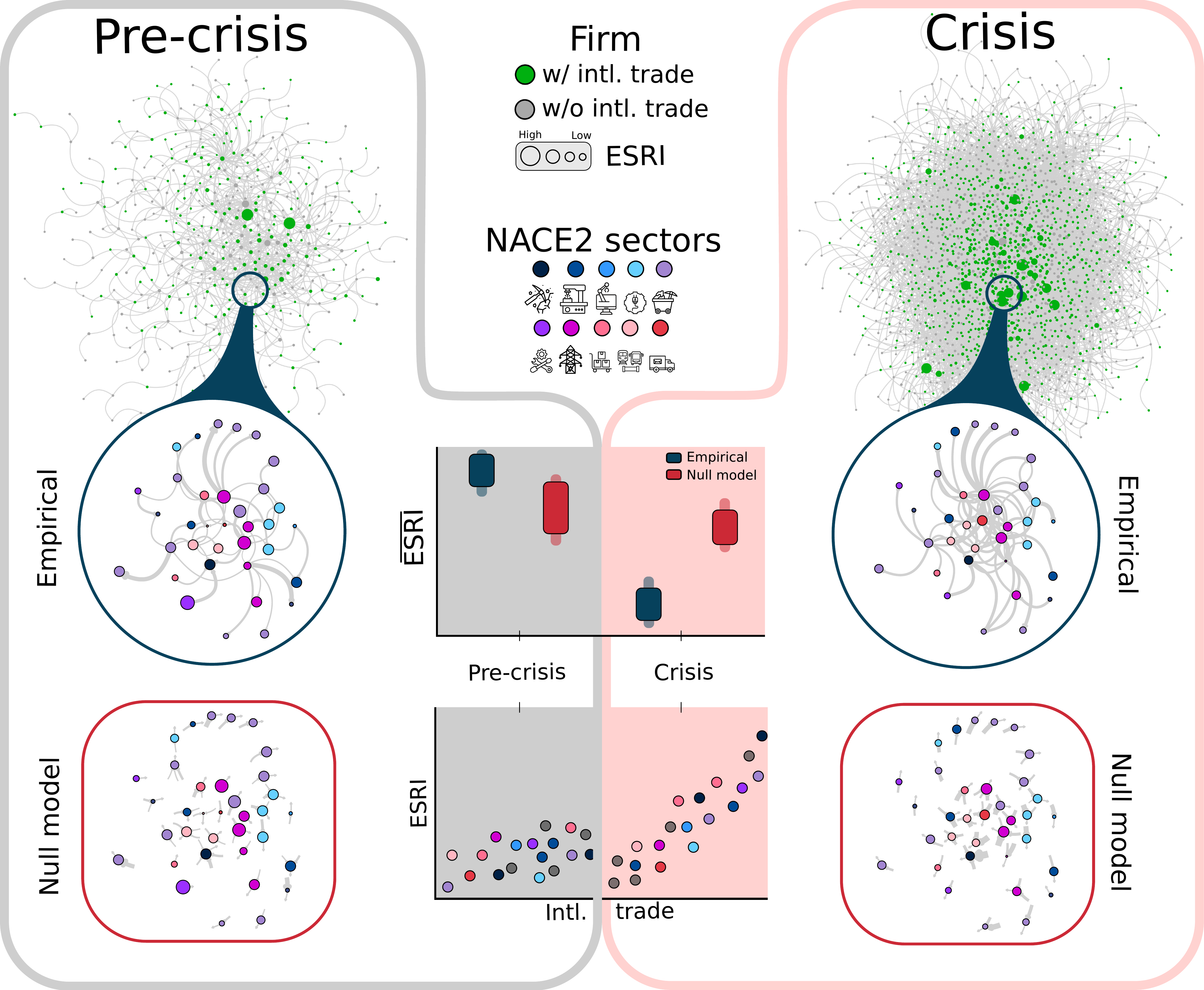}
	\caption{\textbf{Empirical and null model networks are compared before and during the Covid-19 crisis, to study the evolution of economic systemic risk and understand the role of international trade.}
		We start with temporal snapshots of the empirical production network. The top of the figure shows a sample sub-network of 1000 nodes, before and during the crisis. Firms are colored according to the presence (green) or absence (gray) of international trade ties, while nodes' size is proportional to firms' ESRI value.
		We then focus on the firms with the highest systemic risk values, highlighted with a circular magnification from the empirical networks, where nodes' color represents firms' industrial (NACE2) sector.
		Finally, the squares in the third row represent the null models for the same set of risky firms, where we show only the constrained quantities (for each firm, its supply volumes from different sectors and total sales).
		Our framework thus allows studying the temporal evolution of systemic risk, how it deviates from the predictions of the null model, and which characteristics of the firms, such as the presence of international trade, are linked to their risk value (as shown by the sample analysis plots in the middle of the figure).}
	\label{fig0}
\end{figure*}

To focus on how firms' specific position in the evolving production network determines their systemic risk properties, beyond what their supply and demand levels would imply,
we benchmark empirical findings to a null network model that randomizes the supply links of the network but preserves 
the sector-specific input and output trade volumes of each firm. 
We construct this null model using the \emph{stripe-corrected gravity model} (s-GM) \cite{ialongo2022reconstructing}, a heuristic maximum entropy formulation tailored to production networks (see Methods). 
We employ the s-GM to generate an ensemble of synthetic networks on which we compute ESRI scores, serving as null values against which we compare empirical observations. 
Importantly, these null ESRI depend solely on the sector-specific trade volumes of firms, and are therefore independent on the particular topology of the production network  \cite{squartini2011analytical, cimini2019statistical}.
Figure \ref{fig0} illustrates our working pipeline.

By constraining the supply and demand of each firm by sector, the s-GM is consistent with the assumption of fixed production function and input portfolio (here proxied by sectors) commonly used to study shock propagation during crises~\cite{pichler2020production}. 
Indeed, previous research has shown that the s-GM can faithfully reproduce systemic risk levels observed in empirical production networks \cite{fessina2024}. However, we expect null model estimates to deviate from the empirical ESRI measured during the Covid-19 shock, due both to extreme fluctuations 
experienced by the system and to firm-level adaptive responses that the random network model cannot capture. 
We are also particularly interested in understanding the role of firms' international trade activities, which can transmit global shocks into domestic markets \cite{dhyne2021trade, inoue2023disruption}, but can also alter firms' importance within local production networks. Importers, for example, can redistribute foreign products locally \cite{tintelnot2018trade},  or replace local sources with international suppliers \cite{furusawa2017global}, while exporters can attract and coordinate local suppliers \cite{dhyne2021trade}. In the context of the Covid-19 shock, we expect disruptions in global supply chains \cite{guan2020global} to affect local systemic risks through both the propagation of external shocks and the adaptation of domestic networks to changing international trade relationships \cite{javorcik2020reshaping}.

Our results reveal that, at the onset of the Covid-19 pandemic, the set of firms bearing the highest systemic risk undergoes a substantial shift, with firms facilitating economic exchanges emerging as central players in the network -- a structural transformation that is absent in the null model. While systemic risk closely follows null expectations before the pandemic, it diverges significantly afterward, reflecting an adaptive reconfiguration of economic interactions that enhances resilience both in absolute terms and relative to benchmark 
predictions. 
Using regression analysis (see Methods), we find that international trade volume becomes a key predictor of firms’ systemic risk dynamics during the crisis, although global linkages alone cannot fully explain the observed patterns. Imports and exports exert opposing influences on local systemic risk through distinct supply and demand mechanisms, underscoring the complexity of risk propagation in globally interconnected economies.

\section{Results}
\subsection*{The Hungarian production network}

Supplier-customer relations in the Hungarian economy are obtained from firm-level value added tax (VAT) transaction data \cite{borsos2020unfolding,diem2022quantifying} (see Methods). 
Since data contain no information on the type of product exchanged, but only the industrial classification of firms, we follow the typical assumption in related studies that every firm produces only one out of $m$ different products (goods or services) \cite{diem2022quantifying}, which is determined by the firm's industry classification. 
Therefore, we assume that each firm $i$ produces product $p_i$ corresponding to its NACE (Statistical Classification of Economic Activities in the European Community~\cite{ramon2021nace}) classification scheme on the 4-digit level, with a total of $m=615$ product categories. 

The data is then represented as a directed weighted network, where nodes correspond to firms and a link from supplier firm $i$ to buyer firm $j$ is established when data report a purchase transaction from $j$ to $i$, with value $w_{i\to j}$ corresponding to the volume of product type $p_i$ delivered from $i$ to $j$.
We aggregate transactions between firms on a yearly basis to build production networks for the time range 2015-2022. 
These networks are extremely large, with hundreds of thousands of firms and millions of transactions. Moreover they are not coherent in time, as the VAT reporting threshold (the minimum value for a transaction to appear in the data) was lowered in 2018 and completely removed in 2020 (see Supplementary Materials S1 for forfordetails on the various thresholds are how they affect our data). To make these networks computationally manageable and at the same time to remove threshold effects, we apply the highest reporting threshold of 2015 to all networks. Further, we only consider firms whose headquarters are based in Budapest (the region with the highest firm concentration and economic activity) and remove the smallest firms (in terms of number of employees and customers) -- see Methods and Supplementary Materials S1 and S3 for further details. 
In this way, we decrease the size of the networks by a factor ten and obtain yearly coherent snapshots of the {\em local} production network of Budapest (see Table \ref{tab_nodes-edges} in Methods). 

\begin{figure*}[!ht]
	\centering
	\includegraphics[width=\textwidth]{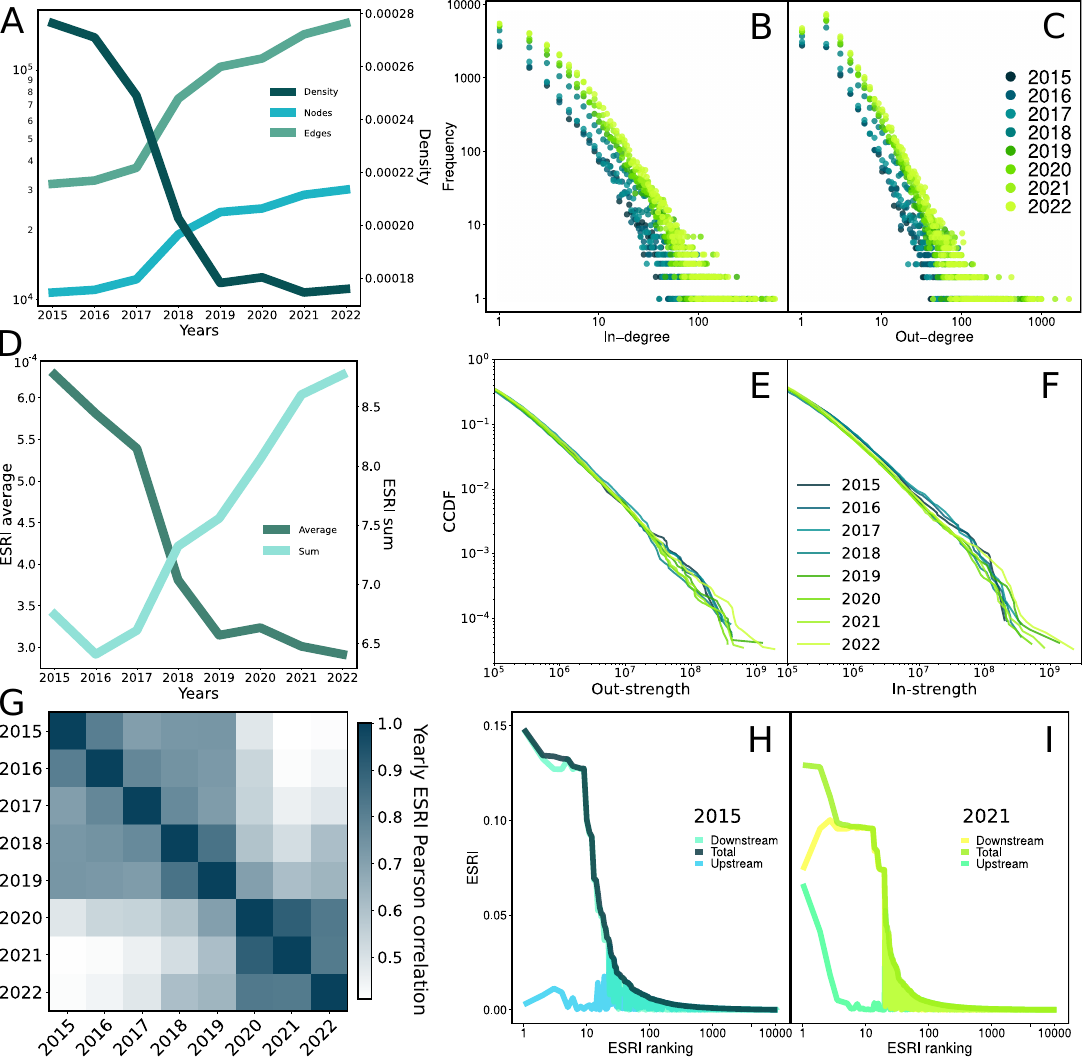}
	\caption{\textbf{Time evolution of topological properties and systemic risk of the empirical production networks.} 
		(A) Number of nodes (firms), links (transactions) and density of the local Budapest production network in the considered time range. (B,C) Frequency distributions of firms' in-degree (number of suppliers) and out-degree (number of customers) for all yearly production networks. 
		(D) Average and total ESRI of firms in the yearly networks. (E,F) Tail of the survival distribution of out-strength values (total sales of firms) and in-strength values (total supply of firms) of yearly production networks. (G) Matrix of Pearson correlation coefficients for ESRI values of firms among different years. (H,I) ESRI values vs rankings for the total, upstream and downstream components for 2015 and for 2021.}
	\label{fig1}
\end{figure*}

The number of firms and transactions in these yearly networks grows substantially over time (Figure \ref{fig1}A). 
Links grow more than the number of nodes $N$ but slower than $N^2$, hence the network density declines.
A big jump is observed in 2018, likely due to the use of online cash registers imposed by the government in 2017\footnote{\url{https://www.vatcalc.com/hungary/hungary-online-real-time-vat-cash-registers/}} and the introduction of the Real-Time Invoice Reporting (RTIR) for all domestic invoices above $100\,000$ HUF in 2018\footnote{\url{https://www.storecove.com/blog/en/e-invoicing-in-hungary/?unbounce_brid=1699967181_5491104_e17e0a18c82a49f0b68fce7880aba2eb}}. 
Thus, from 2018 onward data start to include many 
small-medium enterprises, which mainly have a few connections, causing the drastic drop of link density.
Concerning node-level statistics for each time snapshot, the trading volume of each firm $i$ is described by its out-strength (total sales) $s_i^\text{out}=\sum_{j}w_{i\to j}$ and in-strength (total purchases) $s_i^\text{in} =\sum_{j}w_{j\to i}$, whereas, the firm's connectivity is described by its out-degree (number of customers) $k_i^\text{out}=\sum_{j}a_{i\to j}$ and in-degree (number of suppliers) $k_i^\text{in} =\sum_{j}a_{j\to i}$, where $a_{i\to j}=1$ if $w_{i\to j}>0$ (and 0 otherwise). 
Similarly to previous research in other production networks \cite{bacilieri2022firm}, we find power-law tails (without cut-off \cite{doi:10.1073/pnas.2013825118}) in the distribution of these quantities (Figure \ref{fig1} B-C-E-F), and that the total sales / number of customers is typically much larger than the total supply / number of suppliers, respectively. 
As observed for the density, the degree distributions change mostly because of the introduction of the RTIR, which brought to surface a part of the economy that was hidden beforehand, thus increasing the average degree of firms (see Supplementary Materials S4). 
Strength distributions are instead rather stable in time.

\subsection*{Evolution of Economic Systemic Risk of firms}

To assess how firm-level systemic risk evolves throughout the years, we compute the \emph{Economic Systemic Risk Index} (ESRI) \cite{diem2022quantifying} of all firms in each yearly network. 
In a nutshell, the systemic risk of a firm is computed by removing it from the network and iteratively propagating the upstream (demand) and downstream (supply) shock to connected firms in the production network, which respectively occur when a customer reduces purchases and a supplier reduces provision for a product. The ESRI of the firm, then, corresponds to the overall output reduction of the economy due to its failure (see Methods for more details on the ESRI methodology, 
Supplementary Materials S26 for details on how the simulations were carried out, and Supplementary Materials S5 and S6 for how ESRI relates to firm's degree and strength values).

We first consider the time series of average and total ESRI values of all firms, for each year  (Figure \ref{fig1}D). 
The decreasing trend of the average ESRI is tied to the decreasing network density, which implies relatively less connections and thus fewer shock propagation channels. 
On the other hand, the growing trend of the total ESRI sum mirrors the growth in number of nodes of the yearly networks.
We then focus on the evolution of single-firm ESRI values, to understand whether they are stable or vary over years -- despite firms frequently rewiring their connections \cite{reisch2025rewiring}. 
By computing the Pearson correlation coefficient of firm-level values for each pair of years, we find strong consistency of ESRI over time (Figure \ref{fig1}G), even when the network structure changes considerably from 2017 to 2018. However, the drop in correlation observed from 2019 to 2020 indicates that the Covid-19 pandemic brought about deep network changes that influenced firms' potential impact on the economy (for further insights into yearly ESRI correlations for upstream and downstream values separately, refer to Supplementary Materials S7; for correlation matrices of node centrality measures, see Supplementary Materials S8 -- interestingly, none of them reproduce the structural breaks of ESRI correlations).

As highlighted in previous research \cite{diem2022quantifying, fessina2024, tabachova2024estimating}, plotting ESRI values versus ESRI rankings reveals a plateau composed of the most risky firms, followed by a steep decrease for higher rankings (see Figure \ref{fig1}H for 2015, Figure \ref{fig1}I for 2021 and Supplementary Materials S9 for the other years).
Plateau firms represent the biggest threat to the network: according to ESRI, their failure would result in the loss of about $10\%$ to $15\%$ of the total output of the local Budapest economy. 
We observe that the shape of the plateau, together with the downstream and upstream values, remains consistent in time until 2019, while starting from 2020 the upstream ESRI shows large positive variations that are responsible for the emergence of a step-like structure of the plateau. In particular, the growth in upstream ESRI scores for the plateau firms can be associated with an increase in the firms' in-strength (see the slightly longer tail of the CCDF in Figure \ref{fig2}F).
A possible explanation (supported by the increase in fraction of essential links during the pandemic, see Supplementary Materials S20) is that firms increased their connections with local firms to replace those with foreign countries, restricted due to Covid.

\subsection*{Empirical vs null model values of ESRI}

To benchmark empirical values of ESRI to the null model, using the s-GM method we construct an ensemble of 100 networks for each year in our data (see Methods for the s-GM description, Supplementary Materials S10 showing how the model is able to reproduce the degree and strength statistics of the empirical network and Supplementary Materials S11 for more insight into how link weights are reproduced). We compute the ESRI values of all firms for each of these networks and then average over the ensemble to obtain null yearly values of ESRI for each firm.

\begin{figure*}[!ht]
	\centering
	\includegraphics[width=\textwidth]{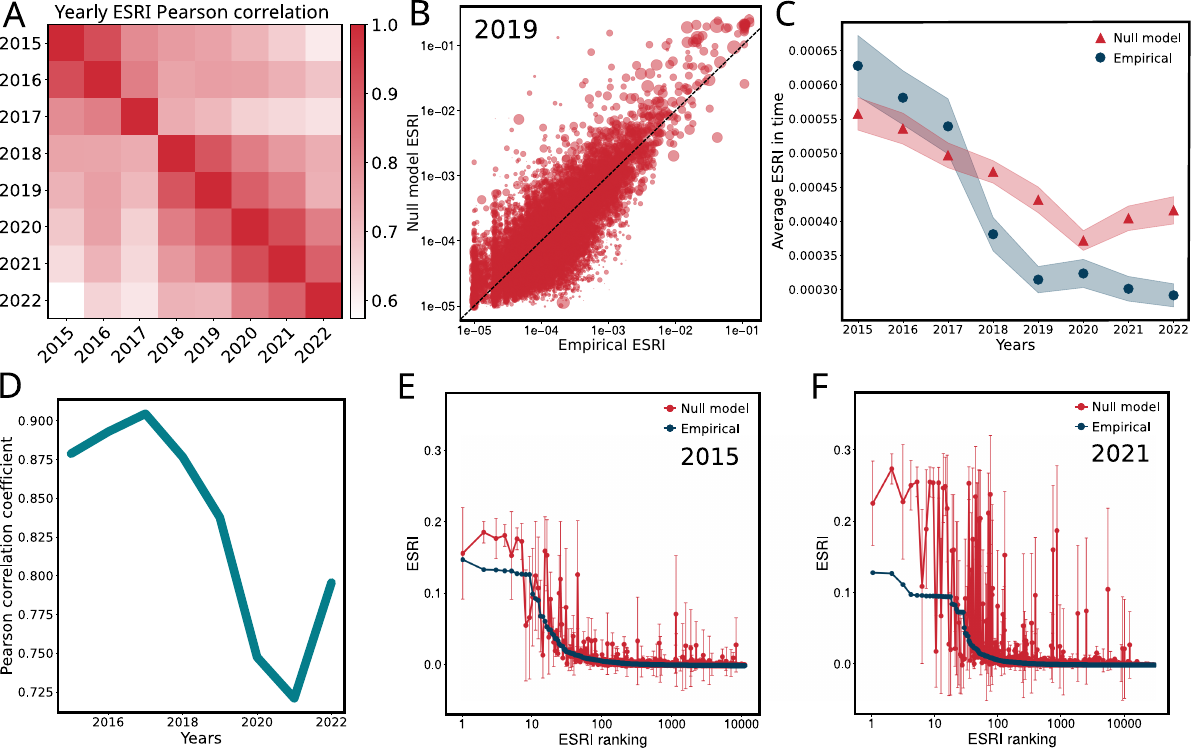}
	\caption{\textbf{Empirical vs null model ESRI values.} 
		(A) Matrix of Pearson correlation
		coefficients for null ESRI values of firms among different years. (B) Empirical vs null ESRI values of individual firms for 2019. The size of each point corresponds to the firm's number of employees, here taken as proxy for size. (C) Mean value of ESRI for empirical and null networks. Shaded area refers to standard deviations (which are larger for the model due to the sampling procedure). (D) Pearson correlation coefficient of empirical vs null ESRI values of individual firms, for each year. (E,F) Empirical and null ESRI ranking for years 2015 and 2021. Error bars represent standard deviations of ensemble values. Model values are ordered according to their empirical ranking.}
	\label{fig2}
\end{figure*}

Similarly to the empirical network, ESRI values in the null model are stable across subsequent years, as revealed by high Pearson correlation coefficients (Figure \ref{fig2}A). However, while for empirical values the highest change is observed between 2019 and 2020 (the onset of Covid), for the null model changes are more pronounced in going from 2017 to 2018, when the real network experienced a considerable growth in nodes and links. This result points out that the null model is not able to fully capture the underlying dynamics of ESRI when the system is affected by exogenous shocks. 
Nonetheless, as shown in Figure \ref{fig2}B for year 2019 (see Supplementary Materials S12 for the other years), there seems to be a good agreement between real and model ESRI values for individual firms \cite{fessina2024}, that is yet not able to reproduce the long-term correlations. In fact we can observe a shift of high-ESRI firms above and of low-ESRI firms below the identity line.
These discrepancies arise as the model slightly overestimates ESRI of the most risky firms (top right corner), while underestimating that of the least risky ones (bottom left corner).
This is mainly due to the sampling method at the basis of the s-GM, for which small firms may happen to get disconnected from the network in some model samples \cite{gabrielli2024critical}. Consequently, these firms have very small ESRI, as they cannot propagate shocks to others. At the same time, since the link density is preserved, bigger firms may then receive slightly more connections, increasing their impact with respect to the empirical case. 

Considering global statistics, the average values of empirical and null model ESRI across all firms are statistically compatible from 2015 to 2017 (Figure \ref{fig2}C). However, from 2018 onward, the model ESRI becomes significantly higher than the real one, possibly because the new firms entering the network in this period follow different market dynamics that are not captured by the null model. 
Notably the deviation increases in 2020-2022, when we witness opposite trends of empirical and null values.
Furthermore, the Pearson correlation coefficient between real and model ESRI scores drops in 2020-2021, during the flare-up of Covid (Figure \ref{fig2}D). These results {reveal} that the null model, which works well both in reconstructing the empirical networks \cite{ialongo2022reconstructing} and reproducing the ESRI dynamics \cite{fessina2024} during normal times, becomes unable to faithfully do so during times of crisis, as the real networks deviate from their null configurations. This deviation may be the result of additional network formation mechanisms at work during that period, which are not included in the null hypothesis underlying the s-GM.

To further corroborate this statement, we compare empirical and model rankings of ESRI for individual firms (Figure \ref{fig2}E-F). 
The overestimation of null ESRI for highly risky firms, discussed above, leads to a higher plateau for model values. 
Overall we observe a good agreement between the empirical and model curves for 2015, with deviations between 0.05 and 0.1 only for a handful of ten firms and much lower for the others -- again confirming that, in normal times, the s-GM provides accurate estimates of single-firm ESRI values \cite{fessina2024}. However, this is not the case for 2021, as the model rankings become more disordered and unable to correctly reproduce not only the plateau, but most of the ranking as well. Indeed we observe deviations above 0.05 (and up to 0.2) for about 40 firms. The mismatch between empirical and model values grows steadily starting from 2018, with 2021 and 2022 yielding the largest deviations (refer to Supplementary Materials S13 for ranking plots of all years, to Supplementary Materials S14 for plots of the deviations between real and null model values, and to Supplementary Materials S15 for distributions of empirical and null model ESRI values).

\begin{figure*}[!ht]
	\centering
	\includegraphics[width=\textwidth]{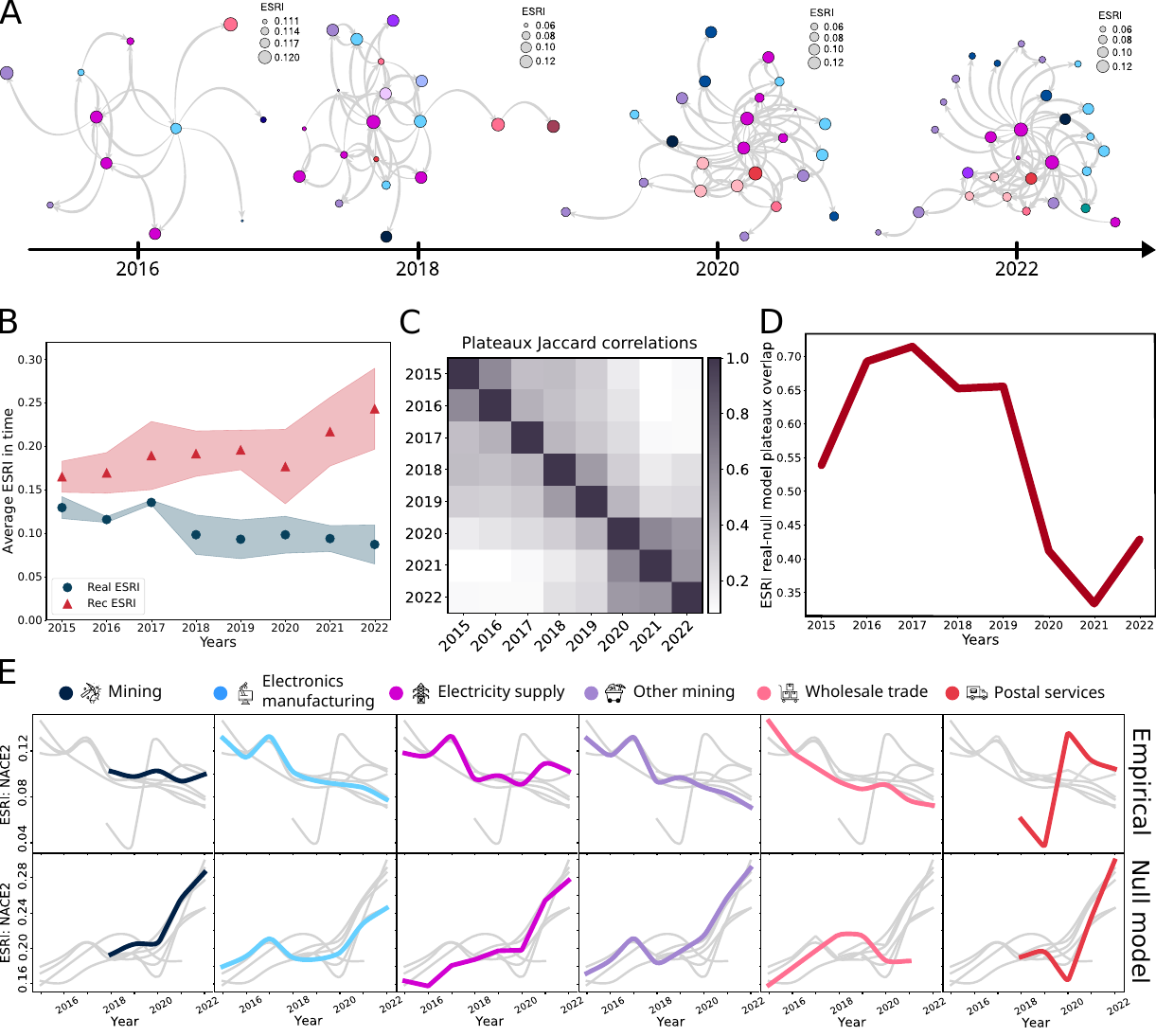}
	\caption{\textbf{ESRI plateaux analysis.} 
		(A) Representation of the empirical network among plateau firms, with node size proportional to ESRI values and color identifying the NACE2 sector. (B) Average ESRI for plateaux firms, displaying the same trend observed for the whole networks. (C) Matrix of Jaccard indices quantifying the overlap among the sets of empirical plateaux firms between different years. (D) Time evolution of the Jaccard coefficient of plateau firms sets between empirical vs model networks in the same year. (E) Evolution of empirical (top panels) and null (bottom panels) ESRI values, for firms appearing more than 3 years in the plateaux, grouped by NACE2 sectors. Each panel highlights the evolution of a single sector (the other sectors are displayed in gray).
		\label{fig3}}
\end{figure*}
\subsection*{Sector composition of ESRI plateaux}

We now focus on the most risky firms, namely those that belong to the ESRI ranking plateaux. For each year, we select plateau firms as the top 0.1\% of the corresponding ESRI ranking (see Supplementary Materials S16 for the full list of NACE4 sectors of these companies).
Figure \ref{fig3}A provides a graphical illustration of the time evolution of the empirical network composed of only the plateau firms, colored by NACE2 sector and sized according to their ESRI values. 
Figure \ref{fig3}B shows the empirical and null model trends of ESRI only for the plateau firms. Notably, these trends are not affected by the 2018 drop that we see for the full networks, and the null model ESRI is always significantly above the empirical value, with the gap between the two widening during Covid. This supports our previous results, as plateau firms are mostly large companies as well as suppliers of essential goods (see Supplementary Materials S7) that are likely present in the network even before the introduction of RTIR in 2018.

Figure \ref{fig3}C further shows that the set of plateau firms changes throughout the years, with a Jaccard index between consecutive years of about 0.5 in ordinary times (meaning that about half of the firms remain in the plateau from one year to the next one) and of 0.7 for the Covid years 2020-2022 (pointing to a rather stable composition of the plateaux).
Among the top firms, a few remain in the plateau for the entire period analyzed. They correspond to NACE2 code 35 (electricity, gas, steam and air conditioning supply) and 28 (manufacture of machinery and equipment). 
In addition, all plateau firms belong to only a handful of industrial sectors (see Supplementary Materials S17 for more details).

We then compare the plateau composition in empirical and null model networks of the same year, where model plateau firms are selected using the corresponding null ESRI rankings. The overlap between the sets of firms in the empirical and model plateaux is high in most years, but decreases noticeably during the pandemic (Figure \ref{fig3}D), when the null model cannot identify most of the most risky firms: in line with the previous result, the model becomes less compatible with the empirical network during Covid-19.
To better analyze the time evolution of the sector composition of the plateau we focus on a stable set of top firms, as those that appear in the empirical plateaux for more than three years. Figure \ref{fig3}E shows the temporal evolution of ESRI for all these firms, summed over NACE2 sectors, for both empirical and null model networks.
Concerning empirical values, we observe a decreasing trend of ESRI for the majority of these sectors, while there is a steep increase in ESRI for the postal activity sector, which represents an emblematic example of the impact of Covid on the production network: due to mobility restrictions, postal services became crucial for the entire system, by allowing trade transactions and maintaining the economy running \cite{gottschalk2023postal}.
The ESRI values for the null model, instead, grow steadily in time, even during Covid years.

We may argue this occurs because the null model is simply driven by the increase in degrees and strengths of the largest firms in the empirical networks in time (see the increasing tail in the distributions of Figures \ref{fig1} B-C-E-F), which results in more interconnections and thus higher systemic-ness for these firms. The growth of these statistics could be partly caused by import bans that forced companies to search for products in the domestic economy, thus increasing the connections we see in the data. 
Since the s-GM directly constrains the strength values, it is already taking into account this increase in local trade, therefore null model ESRI values of large firms become higher during Covid.
On the other hand, in the empirical case, the increase in connections and strengths does not result in increasing ESRI for large firms which, apart from the postal services, display a decreasing trend. In fact, while the postal services represented the only available alternative to bypass lockdown measures, resulting inevitably in an increase of its ESRI value, other firms could decide how to eventually rewire their links (for example to cope with import bans). Notably, this optimization process at the level of individual firms also led to a more resilient network configuration at the global level.
Of course, these mechanisms at work during times of crises are not captured by the null model (see Supplementary Materials S20 for further analyses).

\subsection*{Determinants of firm-level systemic risk}

\begin{table*}[!ht]
	\centering
	\caption{Fixed-effects panel regressions with empirical ESRI as the dependent variable. Given the high correlation between Import and Export, the two variables have been taken into consideration separately. Unreported control variables include Employment. $^{***}$,$^{**}$, $^{*}$ denote significance at the 1\%, 5\%, 10\% level. Standard errors in parentheses.} 
	\begin{tabularx}{\textwidth}{lXXXXXXXX}
		\toprule
		\multirow{2}*{\textbf{}} & \multicolumn{8}{c}{\textbf{ESRI}}\\
		& (1) & (2) & (3)\newline(2015-2019) & (4)\newline(2020-2022) & (5) & (6) & (7)\newline(2015-2019) & (8)\newline(2020-2022)\\
		\midrule
		$ESRI_{model}$ & $0.319^{***}$ \newline (0.004) & $0.319^{***}$ \newline (0.004) & $0.287^{***}$ \newline (0.007) & $0.243^{***}$ \newline (0.007) & $0.319^{***}$ \newline (0.004) & $0.318^{***}$ \newline (0.004) & $0.286^{***}$ \newline (0.007) & $0.242^{***}$ \newline (0.007)\\
		$Import$ & $0.003^{***}$ \newline (0.000) & $0.008^{***}$ \newline (0.002) & $0.005$ \newline (0.003) & $0.020^{***}$ \newline (0.004) &  &  &  & \\
		$Export$ &  &  &  &  & $0.003^{***}$ \newline (0.001) & $0.006^{*}$ \newline (0.003) & $-0.003$ \newline (0.004) & $0.021^{***}$ \newline (0.006)\\
		$Strength_{out}$ & $0.058^{***}$ \newline (0.003) & $0.061^{***}$ \newline (0.003) & $0.063^{***}$ \newline (0.005) & $0.103^{***}$ \newline (0.006) & $0.058^{***}$ \newline (0.003) & $0.061^{***}$ \newline (0.003) & $0.063^{***}$ \newline (0.005) & $0.102^{***}$ \newline (0.006) \\
		$Strength_{in}$ & $0.039^{***}$ \newline (0.001) & $0.038^{***}$ \newline (0.001) & $0.041^{***}$ \newline (0.001) & $0.032^{***}$ \newline (0.001) & $0.039^{***}$ \newline (0.001) & $0.038^{***}$ \newline (0.001) & $0.040^{***}$ \newline (0.001) & $0.032^{***}$ \newline (0.001) \\
		$Essentiality$ & $0.469^{***}$ \newline (0.005) & $0.480^{***}$ \newline (0.005) & $0.480^{***}$ \newline (0.009) & $0.491^{***}$ \newline (0.009) & $0.470^{***}$ \newline (0.005) & $0.474^{***}$ \newline (0.005) & $0.480^{***}$ \newline (0.008) & $0.480^{***}$ \newline (0.008) \\
		\textit{market share} & $-0.034$ \newline (0.042) & $-0.024$ \newline (0.042) & $-0.039$ \newline (0.060) & $0.056$ \newline (0.093) & $-0.033$ \newline (0.042) & $-0.023$ \newline (0.042) & $-0.045$ \newline (0.060) & $0.057$ \newline (0.093) \\
		$Import \times Strength_{out}$ &  & $-0.001^{***}$ \newline (0.001) & $-0.001$ \newline (0.001) & $-0.004{***}$ \newline (0.001) &  &  &  &  \\
		$Import\times Strength_{in}$ & & $0.001^{***}$ \newline (0.000) & $-0.000$ \newline (0.000) &  $0.001^{***}$ \newline (0.000)&  &  &  & \\ 
		$Import \times Essentiality$ &  & $-0.004^{***}$ \newline (0.001) & $-0.003^{*}$ \newline (0.002) & $-0.004^{**}$ \newline (0.002) &  &  &  &  \\
		$Export \times Strength_{out}$ &  &  &  &  &  & $-0.002^{***}$ \newline (0.001) & $0.000$ \newline (0.001) & $-0.006^{***}$ \newline (0.001) \\
		$Export \times Strength_{in}$ &  &  &  &  &  & $0.002^{***}$ \newline (0.000) & $0.001^{***}$ \newline (0.000) & $0.002^{***}$ \newline (0.001) \\
		$Export \times Essentiality$ &  &  &  &  &  & $-0.004^{***}$ \newline (0.001) & $-0.007^{***}$ \newline (0.002) & $0.002$ \newline (0.002) \\
		\midrule
		N & 137956 & 137956 & 65639 & 72317 & 137956 & 137956 & 65639 & 72317\\
		$R^2$ & 0.388 & 0.389 & 0.357 & 0.310 & 0.388 & 0.389 & 0.357 & 0.310\\
		F Statistic & $4085.666^{***}$ & $3369.078^{***}$& $1408.953^{***}$ & $1349.365^{***}$& $4084.467^{***}$ & $3370.141^{***}$ & $1410.481^{***}$ & $1347.189^{***}$\\
		\bottomrule
	\end{tabularx}
	
	\label{tab_regressions}
\end{table*}

To understand the drivers of firm-level systemic risk in the local production network, we estimate the ESRI value of firms in each year $t$ using a multivariate regression, in which explanatory variables include the systemic risk indicator of the s-GM null model ($ESRI_{model}$). To explain the variance in empirical ESRI that null values cannot describe, we add as regressor further company characteristics that are important in the ESRI computation, such as firm size measured by number of employees ($Employment$), its importance in the local trade network measured by in-strength and out-strength of trade flows ($Strength_{in}$, $Strength_{out}$) and $Essentiality$ -- the number of customers for which the firm's input is essential, computed according to the essentiality matrix of \cite{pichler2020production}.
In addition, we include volume of international trade ($Import$, $Export$) that, even if not directly used in the ESRI computation, we expect to be non-trivially related to systemic risk in the local economy -- by generating spillovers from international product flows through local links in the production network. The regression includes company fixed-effects that capture further unobserved firm characteristics and year dummies that control for yearly fluctuations (see Methods for a more detailed presentation of the regression model and the variables included, and Supplementary Materials S18 for the correlation table of firm features).

Table \ref{tab_regressions} shows a series of regressions, in which models including either $Import$ (Models 1-4) or $Export$ (Models 5-8) are separated due to the high correlation of these variables. Regressions (1) and (5) allow to understand the influence each variable has on the value of the empirical ESRI for a given firm. As expected from the previous analyses, we find a strong correlation between empirical ESRI and null model values. Growing intensity of local trade is associated with increasing ESRI, especially if the company's supply is essential for the customers' businesses. 
Size effects are stronger for out-strength than in-strength, because the former is related to downstream shocks, which propagate with a generalized Leontief production function, while upstream shocks propagate linearly. $Essentiality$ has a strong positive correlation with $ESRI$, signaling that systemic risk is higher for those firms that provide essential inputs for many other firms in the local economy.
Overall, the contributions of these variables containing information of the local trade network are in line with expectations and also seem to capture the scale effects, since $Employment$ has no significant relationship with ESRI (not reported in \ref{tab_regressions}).

The positive and significant coefficients of $Import$ (1) and $Export$ (5) signal that international trade can also increase systemic risk of firms in national economies. 
This result is non-trivial, but paired with local links, increasing international trade can ripple through to local economic systems. 
To disentangle these mechanisms, we add the interaction terms between international trade and local trade variables and find that the impact of international trade on local systemic risk through domestic links (represented by in- and out-strengths) shows a mixed picture.
On the one hand, the negative interaction term in (2) and (4) between $Import$ and $Strenght_{out}$ as well as $Essentiality$ signals that import does not have a spillover impact on systemic risk in the local economy of Budapest. 
Due to the dominance of multinational companies in manufacturing sectors that are more embedded in global value chains than in local economies \cite{halpern2015imported, juhasz2024colocation}, importing companies probably sell their products predominantly on export markets, instead of distributing imported goods through local links. As a result, we see a negative association of $Import \times Strength_{out}$ with $ESRI$. However, the positive interaction term between $Import$ and $Strength_{in}$ in Models (2) and (4) signals that firms can also complement domestic supplies and thus increase the firm-level $ESRI$ altogether. 
On the other hand, the interaction of $Export$ with $Strength_{out}$ and $Essentiality$ is negatively correlated with $ESRI$ (6), suggesting that increasing exports can decrease systemic risk in the domestic economy, in case they substitute local revenues. Yet, the interaction of $Export$ with $Strength_{in}$ has a positive relationship with $ESRI$, indicating that growing export can increase systemic risk of firms if entering international markets is paired with growing local supplies. We further find that a firm's market share is never significant, a surprising result only at first glace, since the market share of firms is reproduced by the null model, thus its effect is already incorporated in the coefficient of $ESRI_{model}$.

Splitting the period into pre-Covid years (3,7) and Covid years (4,8) reveals that both $Import$ and $Export$ become significantly correlated with ESRI only during Covid. This can be explained considering that during the Covid-19 pandemic there was an overall drop in trade due to import and export difficulties all over the globe. Thus, only the most important firms were able to maintain trade with foreign countries, and these firms get the highest ESRI values (see Supplementary Materials S19 for a correlation of ESRI, essentiality and size, proxied by the firm's out-strength, Supplementary Materials S20 for the yearly evolution of fraction of essential links, and Supplementary Materials S21 and S25 for results obtained with a weighted version of the essentiality score). 

Consequently, the above reported impact of import on national systemic risk through local production networks is only apparent during the Covid period in our case (4). However, we find that the effect of export is slightly different with respect to Covid. While exports as supplements for domestic revenues only decrease ESRI during Covid, they only substitute essential links in the pre-Covid period. We find a stable positive and significant joint effect between $Export$ and $Strength_{in}$ in both pre- and Covid years. This signals that increasing export may increase systemic risk in national production networks if it increases the demand for local inputs.

We run a series of robustness checks that provide further support for the above results. Pooled OLS regressions with year fixed-effects and cross-sectional linear regressions run by years are reported in Supplementary Materials S22, S23 and S24.

\section*{Conclusions}

Understanding how firm-level systemic risk evolves and reacts to external disruptions is key to explaining shock propagation in economic networks. We analyzed the temporal evolution of the Economic Systemic Risk Index (ESRI) \cite{diem2022quantifying} for firms in the Budapest VAT transaction network (2015–2022) and compared empirical patterns with the stripe-corrected gravity model \cite{ialongo2022reconstructing}, a null model that constrains the input by sector and output volumes of each firm in the economy. 
In normal times, ESRI is highly persistent, reflecting stable structural regularities. This persistence breaks sharply during Covid-19 (2020–2022), when systemic risk declines both absolutely and relative to the null benchmark. This decorrelation shows that systemic importance is not a fixed firm attribute but an emergent, context-dependent property of the production network.

We interpret the observed ESRI reduction during the pandemic as evidence of an adaptive behavior: when faced with factory closures, mobility restrictions, and disruptions to international trade, firms actively rewire their supply-chain connections, leading to a global network configuration that is more resilient than predicted by the null model. This result suggests that decentralized firm-level optimization, even in the absence of coordination or planning, can give rise to collective resilience. More broadly, our findings support a view of the economy as a complex adaptive network rather than a system that fluctuates around a fixed equilibrium.
Sectoral heterogeneity further supports this dynamic view. While most initially high-risk firms experienced declining ESRI, the postal sector became increasingly systemically relevant during the pandemic, acting as a connectivity enabler. This shift, not captured by the null model, highlights the need for crisis policies that protect network coordinators and logistics providers, not only large producers.

Regression results reveal that international trade is strongly associated with ESRI, but only during the Covid-19 pandemic. This finding signals an effect of trade links on systemic risk in local production networks during crises. However, our evidence on mechanisms remain asymmetric. On the one hand, we  find that export increases systemic risks more if companies develop their local supply connections as well, suggesting a link between global value chains and local systemic risk. On the other hand, we find no evidence for the propagation of international effects in the local system. However, imports increase systemic risks more for those companies that also increase their local inputs, suggesting that combining imported and local intermediary inputs can generate channels of shock propagation on the local level.

Overall, this work paves the way for a broader temporal analysis of systemic risk and shock propagation in economic networks, emphasizing the importance of adaptive behavior and contextual systemic relevance. Maximum-entropy models provide a useful normal-times benchmark \cite{fessina2024}, but persistent deviations during crises signal they cannot reproduce endogenous adjustments \cite{squartini2013early, Gualdi:2016aa,ferracci2022systemicriskinterbanknetworks}. In particular, the s-GM we use resembles a thermodynamic construction of networks at the Walrasian economic equilibrium \cite{bargigli2016statistical,bardoscia2021physics}, according to which agents in an exchange economy care only about final allocations and are indifferent with respect to various market configurations (i.e., network structures) that realize the same allocations.
\footnote{This parallel holds by assuming that constraints are imposed, not solved for (as endogenous outcomes emerging from technologies, preferences, and market clearing), that there are no agents, prices, supply and demand, nor conditions on costs, profits, efficiency.}
Therefore, comparison with the null model allows us to assess how far the empirical networks are from their thermodynamic equilibrium configurations induced by the production structure of each firm, and whether significant deviations appear in the presence of exogenous macroeconomic shocks (for a detailed discussion see Supplementary Materials S2).

We stress that the time span of our analysis covers years with different values of the reporting threshold. 
Our systematic filtering procedure (see Methods) is specifically meant to remove effects of the changing threshold and highlight other structural changes -- in particular, the introduction of RTIR in 2017/2018 and the adaptive reconfigurations due to the Covid-19 pandemic in 2020. 
Of course, there could be confounding factors affecting our results, since the threshold change and these events both took place during the same years. 
Therefore, it may be possible that the filtering procedure cannot offset all the changes in network topology the threshold change is causing. 
Further analyses (see Supplementary Materials S1, S8 and S20 support the robustness of our filters and point to a changing link formation mechanism taking place during Covid. Nevertheless, a robust comparative analysis would ideally require a dataset that is homogeneous from the outset, with consistent reporting criteria throughout the selected time frame.

Another key limitation of our study is the computational intensity of ESRI, which restricted the analysis to Budapest and a limited null-model ensemble. 
Additionally, although Hungary faced lockdown restrictions similar to those of other European countries, a wider generalization of our results to other nations is not trivial. 
Given data availability, future work may extend the framework to broader contexts and incorporate measures of firm resilience and spatial exposure to better understand shock absorption and recovery. Further research is needed to get actionable insights on what the firms and economic policies could do to reduce systemic risk.

\section*{Methods}

\subsection*{Data description and filtering}\label{data_descr}

The Hungarian production network is obtained from VAT transaction data as follows. 
For every sales transaction between a supplier $i$ and buyer $j$, we have the monetary value of the goods and services sold, $V_{ji}$ in forints (HUF), and the tax amount paid $T_{ji}$. We use $V_{ji}$ as an estimate for the volume $w_{ij}$ of product type $p_i$ delivered from $i$ to $j$. 
We aggregate all transactions on a yearly basis by summing all exchanges between two firms that take place in a given year. 

Transactions for years 2015 to 2017 appear in the dataset only if they exceed a reporting VAT threshold of $t = 1\,000\,000$ HUF (about $2\,500$ EUR), which was lowered in 2018 and completely removed in 2020.
In order to make the whole dataset homogeneous we apply the same threshold of 2015 for all other years, thus removing transactions with value less than $t$. We also implement two additional filters. 
To build a local production network and remove possible geographical effects or biases, we only consider firms whose headquarters are based in Budapest (which represent 29\% of the total firms in Hungary).
Furthermore, in order to make the yearly networks easier to handle, for each snapshot we also remove the {smallest firms, selected as those having} a number of employees $\le$ 11 and also a number of customers $k^\text{out}\le 2$ (see Supplementary Materials S1 and S3 for further information on the goodness of the filters).

\begin{table}[!ht]
	\centering
	\begin{tabular}{ccccc}
		\toprule
		\multirow{2}*{\textbf{Year}} & \multicolumn{2}{c}{\textbf{Pre-filtering}} & \multicolumn{2}{c}{\textbf{Post-filtering}}\\
		& Firms & Transactions & Firms & Transactions\\
		\midrule
		2015 & 88992 & 221200 & 10745 & 31893\\
		2016 & 93258 & 225165 & 11032 & 32967\\
		2017 & 101590 & 254940 & 12267 & 37456\\
		2018 & 272410 & 1373207 & 19224 & 74866\\
		2019 & 305492 & 2059287 & 24039 & 102928\\
		2020 & 440715 & 12457336 & 24921 & 111953\\
		2021 & 448750 & 18147540 & 28580 & 159308\\
		2022 & 457239 & 19050904 & 30098 & 159308\\
		\bottomrule
	\end{tabular}
	\caption{Number of firms and transactions for all yearly networks, before and after the filtering procedure.} 
	\label{tab_nodes-edges}
\end{table}

\subsection*{The Economic Systemic Risk Index}

The \emph{Economic Systemic Risk Index} (ESRI) quantifies the systemicness of a firm by evaluating the output reduction experienced by the whole production network due to its failure. 
More in detail, the removal of a firm from the network (meaning that it stops buying from its suppliers and supplying to its customers) generates an upstream and a downstream shock, which are propagated to all other firms by recursively updating their production levels. 
The downstream shock propagates according to a generalized Leontief {\em production function} \cite{diem2022quantifying}. Inputs from \emph{essential} sectors set a Leontief-kind of constraint on the output of firms, while the inputs from \emph{non-essential} sectors are treated in a linear way. The identification of essential products, according to their NACE4 industry classification, is provided by expert based surveys \cite{pichler2020production,pichler2021and}, while the technical coefficients of the production function are calibrated on the empirical network.
The shock propagation dynamics is iterated until convergence, when the final production level of every firm is used to compute the ESRI of the initially defaulted company.  
We remand the reader to \cite{diem2022quantifying} for full details of the method, and Supplementary S26 for the description of the specific setting we employ. Below is a brief description of the algorithm.

To represent the initial default of firm $j$, while other firms remain unaffected, the initial shock is set as $\psi_j=0$ and $\psi_i=1$ $\forall i\neq j$. 
Then, ESRI dynamics is based on: 
i) the downstream impact matrix, whose element $\Lambda_{ji}^\text{d}$ determines the fraction of production firm $i$ loses if firm $j$ stops supplying to it: 
$\Lambda_{ji}^\text{d} = w_{j\to i}/s_{g_j\to i}$ if $g_j \in \text{Ess}_i$ and $\Lambda_{ji}^\text{d} = w_{j\to i}/\sum_{g'}s_{g'\to i}$ otherwise, 
where $\text{Ess}_i$ is the set of products that are essential for firm $i$; ii) the upstream impact matrix, whose element $\Lambda^\text{u}_{ji}= w_{i\to j}/s_i^{out}$ determines the fraction of production firm $i$ loses if firm $j$ stops buying from it.

After applying the initial shock, downstream and upstream shocks are propagated to any firm $i$ through two iterative equations:
\begin{equation} \label{eq:esri_downstream}
	h_i^{\text{d}}(n+1)   =   \min\Big[
	\min_{g \in \text{Ess}_i} \Big(\tilde{\Pi}_{ig}(n)   \Big), \;
	\tilde{\Pi}_{i}(n)
	, \psi_i \Big]
\end{equation}
\begin{equation} \label{eq:esri_upstream}
	h^u_{i}(n+1) = \min \Big[ \sum_j \Lambda^\text{u}_{ji}h^\text{u}_j(n), \psi_i  \Big]  
\end{equation}
Here $n$ indicates the time step of the propagation, while $h_{i}^{d}(n)$ and $h_{i}^{u}(n)$ are the residual fraction of production of firm $i$ at step $n$ following the propagation of the downstream and upstream shock, respectively, with initial values $h_i^d(0)=h_i^u(0)=\psi_i$. 
In the equations above, the relative amount of essential inputs in sector $p\in\text{Ess}_i$ available for firm $g$ is $\tilde{\Pi}_{ig}(n) = 1 - \sum_{j\in g} \sigma_j(n) \Lambda^{d}_{ji} \big(1-h_j^\text{d}(n) \big)$, 
the relative share of all non-essential inputs is $\tilde{\Pi}_{i}(n) = 1 - \sum_{g\notin \text{Ess}_i} \sum_{j\in g} \sigma_j(n) \Lambda^{d}_{ji} \big(1-h_j^\text{d}(n) \big)$, while the firms' market share, used as a proxy for how replaceable it is for its customers, reads 
$\sigma_j(n) =  \min \left[1, s_j^{out}/(\sum_{l\in g_j} s_l^{out} \,h_l^d(n)) \right]$. 
After the two, independent shocks of eqs. \eqref{eq:esri_downstream} and \eqref{eq:esri_upstream} have converged at $n=n^*$, the residual fraction of output of firm $i$ is computed as $h_i(n^*)=\min\{h_i^{d}(n^*),h_i^{u}(n^*)\}$.
Finally, the ESRI of firm $i$ is computed by summing the output reduction of each firm in the network, stemming from its initial default:
\begin{equation}
	ESRI_i = \sum_{j} \frac{s_j^{out}}{\sum_{l} s_l^{out}} [1 - h_j(n^*)]
\end{equation}

\subsection*{The Stripe-Corrected Gravity Model}

Unbiased null models of networks can be constructed using constrained entropy maximization, which generates an ensemble of networks that are maximally random but preserve (constrain) selected features of the empirical system \cite{jaynes1957information}.
In physics jargon, these networks are in ``thermodynamic equilibrium'' with the imposed constraints. 
This construction thus defines a rigorous null hypothesis that the constrained features can explain any observation on the network \cite{squartini2011analytical,cimini2019statistical}. 
This validation approach is widely used in economic and financial network studies \cite{fagiolo_null,squartini2013early,Gualdi:2016aa,ferracci2022systemicriskinterbanknetworks,fessina2024pattern,zelbi2025mitigation}.

In order to define unbiased null models of sparse weighted networks (which is the case of production networks), the typical recipe is to constrain node strengths (the total weight of incident connections) as well as node degrees (the number of connections per node)  \cite{mastrandrea2014enhanced,gabrielli2019}. This can be done using a heuristic two-step procedure, known as \emph{density-corrected gravity model} (d-GM), assessing first whether two nodes are connected and then the value of the link weights \cite{cimini2015systemic, parisi2020horse}. 
The first step is based on the canonical \emph{configuration model} \cite{park2004statistical}, where model parameters controlling the degrees can be replaced by node strengths using a \emph{fitness ansatz} that connectivity is proportional to size \cite{caldarelli2002scale,garlaschelli2004fitness}. 
The \emph{stripe-corrected Gravity Model} (s-GM) \cite{ialongo2022reconstructing} improves the d-GM recipe by constraining the sector-specific strength of each node (rather than the overall strength), preserving in this way the input-output productive structure of each firm. 
Specifically the model preserves, for each firm $j$, the total output $s_j^{out}=\sum_{i} w_{j\to i}$ as well as the list of input quantities by industry $g$, namely $s_{g \to  j} = \sum_{i \in g} w_{i\to j}$, which is a good proxy for the amount of products by industry that the firm uses to produce its output.

Quantitatively, the two-step procedure of the s-GM consists in first estimating the probability of a connection from supplier $i$ to customer $j$:
\begin{equation}
	p_{i\to j} = \frac{z_{g_i} s_i^{out} s_{g_i \to j}}{1 + z_{g_i} s_i^{out} s_{g_i \to j}}, \quad \forall i \neq j
\end{equation}
where $g_i$ is the industrial sector of the supplier $i$, and then assigning the link weight according to:
\begin{equation}
	w_{i\to j} = \frac{s_i^{out} s_{g_i \to j}}{w_{g_i}^{tot} p_{i\to j}}, \quad \forall i \neq j
\end{equation}
where $w_{g_i}^{tot} = \sum_{k \in g_i} \sum_{j} w_{k\to j} = \sum_{k \in g_i} s_k^{out} = \sum_{j} s_{g_i \to j}$ 
is the total outgoing strength of sector $g_i$.
We employ the multi-z variant of the s-GM, thus we have a different parameter $z_{g_i}$ for each industrial sector, which can be determined by solving the equations
%\begin{equation}
$\sum_{k \in g_i}\sum_{j \neq k} p_{k\to j} = L_{g_i}$
%\end{equation}
where $L_{g_i} = \sum_{k \in g_i}\sum_{j} a_{k\to j}$
is the number of outgoing links from sector $g_i$.
Thus by construction the model preserves, on average, the total number of links as well as the out-strength and the in-strength by sector of each firm, reproducing the production structure of the empirical network:
\begin{equation}
	\begin{split}
		\langle s_i^{out} \rangle = \sum_{j} \langle w_{i\to j} \rangle = s_i^{out} \frac{\sum_{j} s_{g_i \rightarrow j}}{w_{g_i}^{tot}} = s_i^{out}, \quad \forall i, \\
		\langle s_{g_i \rightarrow j} \rangle = \sum_{k \in g_i} \langle w_{k\to j} \rangle = s_{g_i \rightarrow j} \frac{\sum_{k \in g_i} s_{k}^{out}}{w_{g_i}^{tot}} = s_{g_i \rightarrow j}, \quad \forall j, g_{i}
	\end{split}
\end{equation}

\subsection*{Regression framework}

To analyze the dynamics of firms' systemic risk, we estimate $ESRI$ values of firm $j$ at year $t$ in a multivariate fixed-effect panel regression, following the formula:
\begin{equation}
	\begin{split}
		ESRI_{j,t} = \alpha + \beta_1 (ESRI_{model})_{j,t} + \beta_2E_{j,t} + \beta_3LT_{j,t} + \\\beta_4IT_{j,t} + \beta_5(LT_{j,t} \times IT_{j,t}) + \mu_j + \epsilon_{j,t}, 
	\end{split}
\end{equation}
where $ESRI_{model}$ denotes the average of ESRI values from the null model, $E_j$ stands for firm size (number of employees), $LT_j$ is the characteristics of local trade measured in the production network (out-strength, in-strength, essentiality score), $IT_j$ denotes the value of international trade (import and export), $\mu_j$ is the firm fixed-effect and $\epsilon$ is the error term. This framework allows for assessing the determinants that can explain firms' systemic risk, beyond the null model expectations. The interaction term between $LT$ and $IT$ enables us investigate how international trade is related to local systemic risk, mediated by local trade channels.

\section*{Acknowledgements}
This manuscript was prepared using the datasets of the Hungarian Central Statistical Office on firm-to-firm VAT transactions and firm import-export. The calculations contained in the document and the conclusions drawn from them are exclusively the intellectual products of Anna Mancini, Balázs Lengyel, Riccardo Di Clemente and Giulio Cimini as the authors. Images for NACE2 sectors appearing in figures were created with Canva Pro.

\section*{Funding}
A.M. and G.C. acknowledge financial support from the National Recovery and Resilience Plan (NRRP), Mission 4 Component 2 Investment 1.1, funded by the European Union - NextGenerationEU: 
Call for tender No. 104 of 02/02/2022 by the Italian Ministry of University and Research (MUR), Project Title: \emph{RENet - Reconstructing economic networks: from physics to machine learning and back}, Concession Decree No. 957 of 30/06/2023, Project code 2022MTBB22 - CUP E53D23001770006;
Call for tender No. 1409 of 14/09/2022 by the Italian Ministry of University and Research (MUR), Project Title: \emph{C2T - From Crises to Theory: towards a science of resilience and recovery for economic and financial systems}, Concession Decree No. 1381 of 01/09/2023, Project code P2022E93B8 - CUP E53D23018320001.\\
B.L. acknowledges financial support from the Hungarian National Scientific
Fund (OTKA K 138970).\\
R.D.C. acknowledges support from the Lagrange Project of the ISI Foundation funded by CRT Foundation.\\

\section*{Author contributions statement}

Study conception and design: BL, RDC, GC. Data analysis: AM, BL. Contributed methods: AM, GC. Discussion and interpretation of results: AM, BL, RDC, GC. Manuscript preparation: AM, BL, RDC, GC.

\section*{Competing interests}

The authors declare no competing interests.

\section*{Preprints}
A preprint of this article can be found at\\ \url{https://arxiv.org/abs/2506.21426}.

\section*{Data availability}
This manuscript was prepared using the datasets of the Hungarian Central Statistical Office. These datasets are not publicly available and can only be accessed after authorization by the institution and after signing a declaration to ensure data protection. They are available in-loco at the DataBank of KRTK-KTI, 1097 Budapest, Tóth Kálmán utca 4. Access to the data is organized through the Head of the Data Bank, Melinda Tir (ELTE Center for Economic and Regional Studies; tir.melinda@krtk.hu), who also provides the necessary infrastructure; however, the legal decision on granting access rests with the Hungarian Central Statistical Office.\\
Sample codes and datasets used to produce our results are available at \url{https://github.com/mnlknt/Evolution-and-reconstruction-of-firm-level-systemic-risk}..

\bibliographystyle{plain}
\bibliography{reference}



\section*{Supplementary material}

# Evolution and determinants of firm-level systemic risk in local production networks

Anna Mancini, Balázs Lengyel, Riccardo Di Clemente, Giulio Cimini

# Supplementary Materials

## S1 Filtering procedure

Figure S5A shows that the majority of firms' headquarters are based in Budapest. Figure S5B instead shows the number of employees versus the firm's ESRI value before applying the last filter on the number of employees and the degree. The two variables are very correlated, thus firms with fewer employees also show lower ESRI scores. All firms that satisfy the condition on employees ($\leq 11$) have a low value of systemic risk.

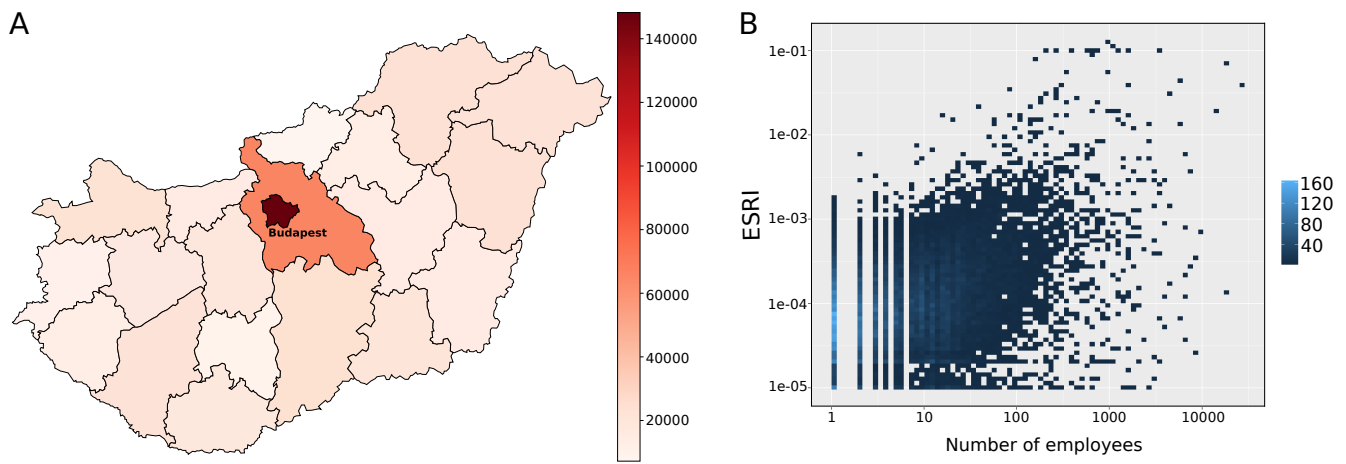

**Figure S5. Filtering procedures.** A) Hungarian NUTS3 counties colored by the number of firms that are based in each one, as of 2019. B) ESRI of each firm versus number of employees.

## S2 NUTS3 aggregation of firms' sales

To proxy the GPD of the different Hungarian regions, we aggregate all transactions between firms on a NUTS3 level, resulting in a network where nodes are NUTS3 regions and links are all monetary transactions between them. Figure S6 shows the out-strength (total sales) for each region, where we can see how the Budapest area is responsible for around 48% of the total out-strength of all regions.

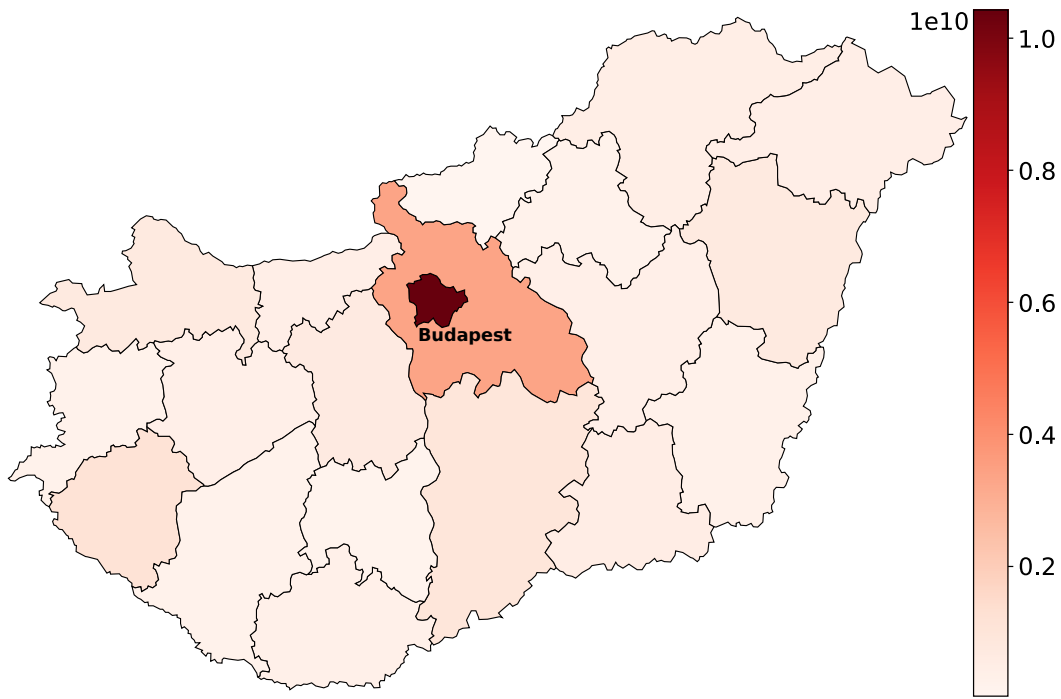

**Figure S6. NUTS3 out-strength.** Out-strength aggregated on a NUTS3 level, for all Hungarian regions.

## S3 ESRI vs degree values

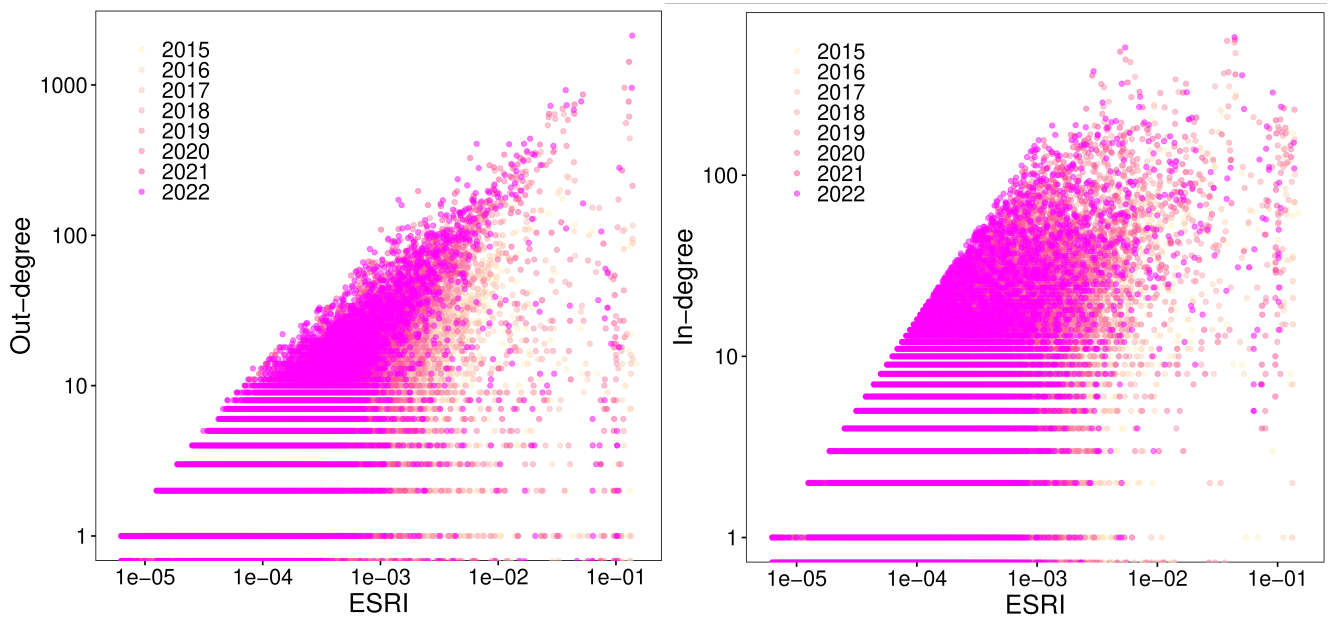

**Figure S7.** Scatter plot of ESRI values vs firm's in- and out-degree.

## S4 ESRI vs strengths

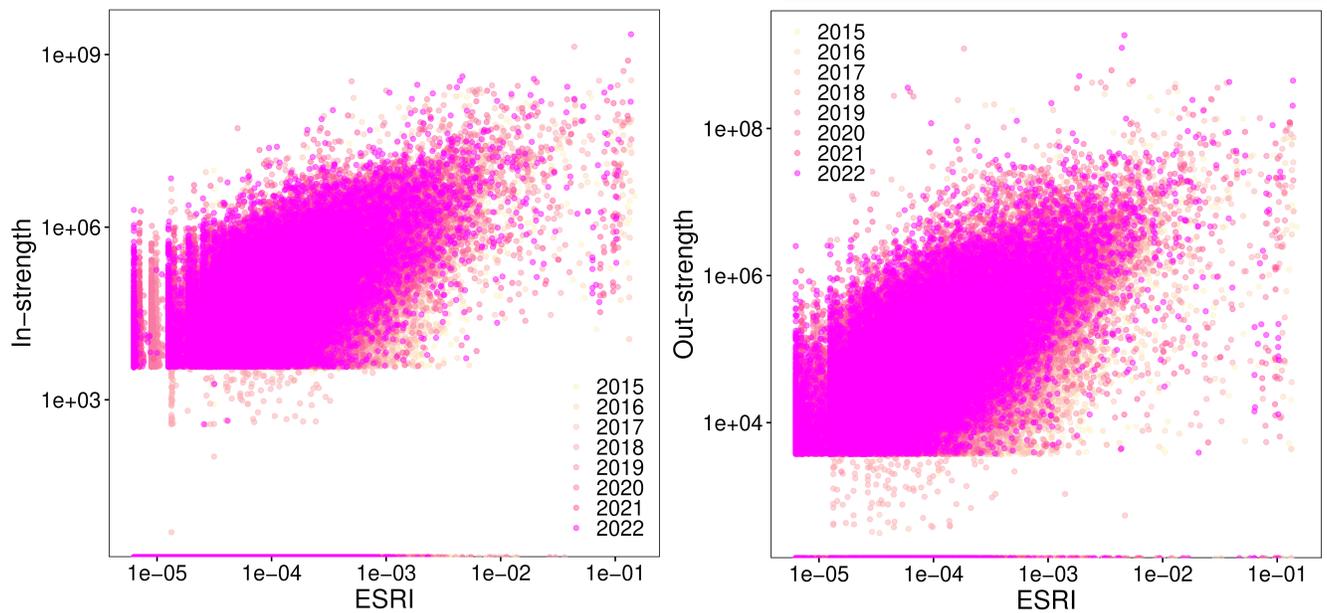

**Figure S8.** Scatter plot of ESRI values vs firm's in- and out-strengths.

## S5 Correlation matrices of ESRI values

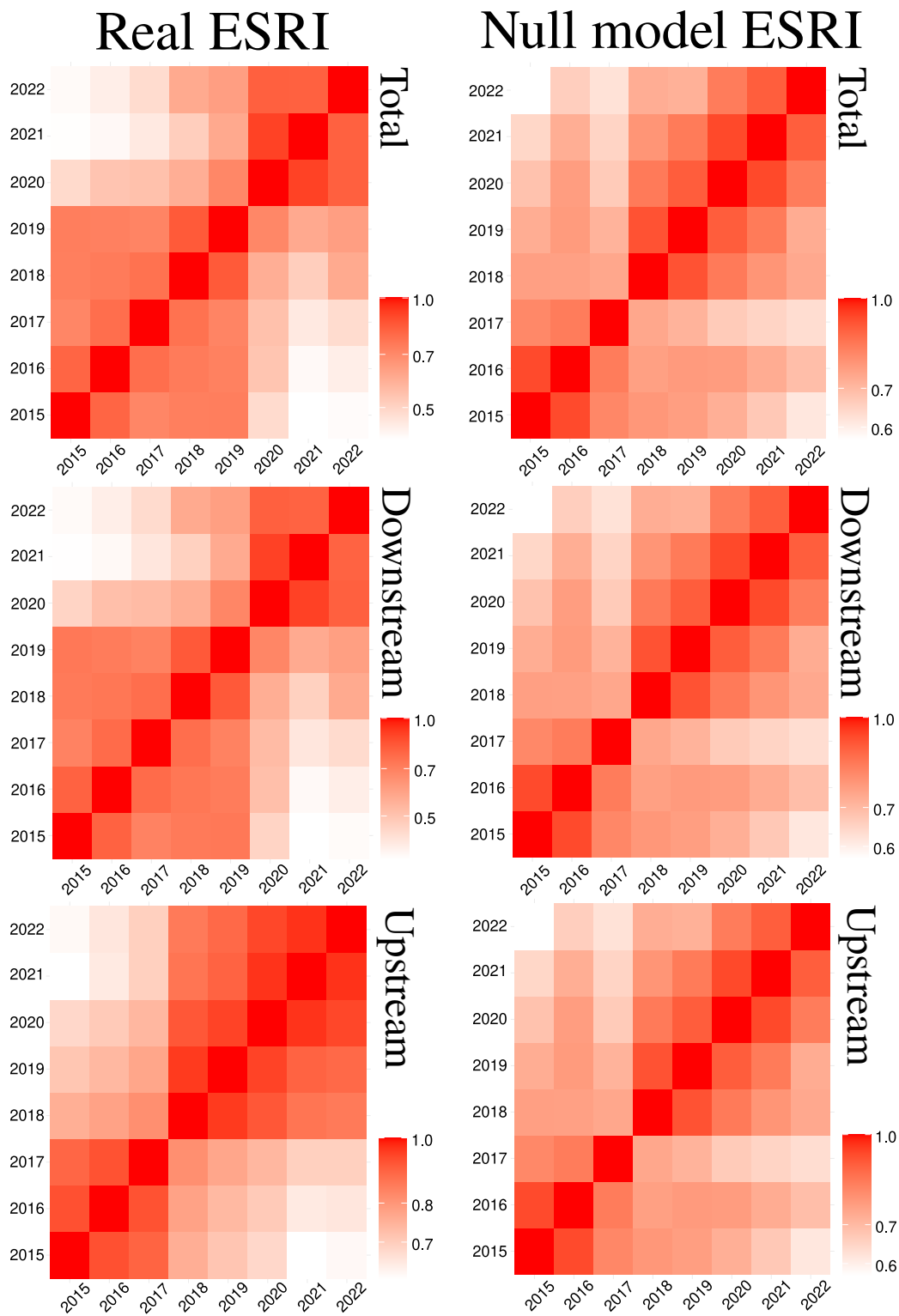

**Figure S9.** Correlation matrices of ESRI values for all years considered. Left column for empirical networks and right column for null models. First row shows matrices for the total ESRI values, second row for the downstream values and third row for the upstream values. It is worth noting how the upstream ESRI values show a disruption in 2018, hinting to a possible change in the structure of the network affecting mostly the in-strengths.

# S6 ESRI empirical rankings for 2015-2022

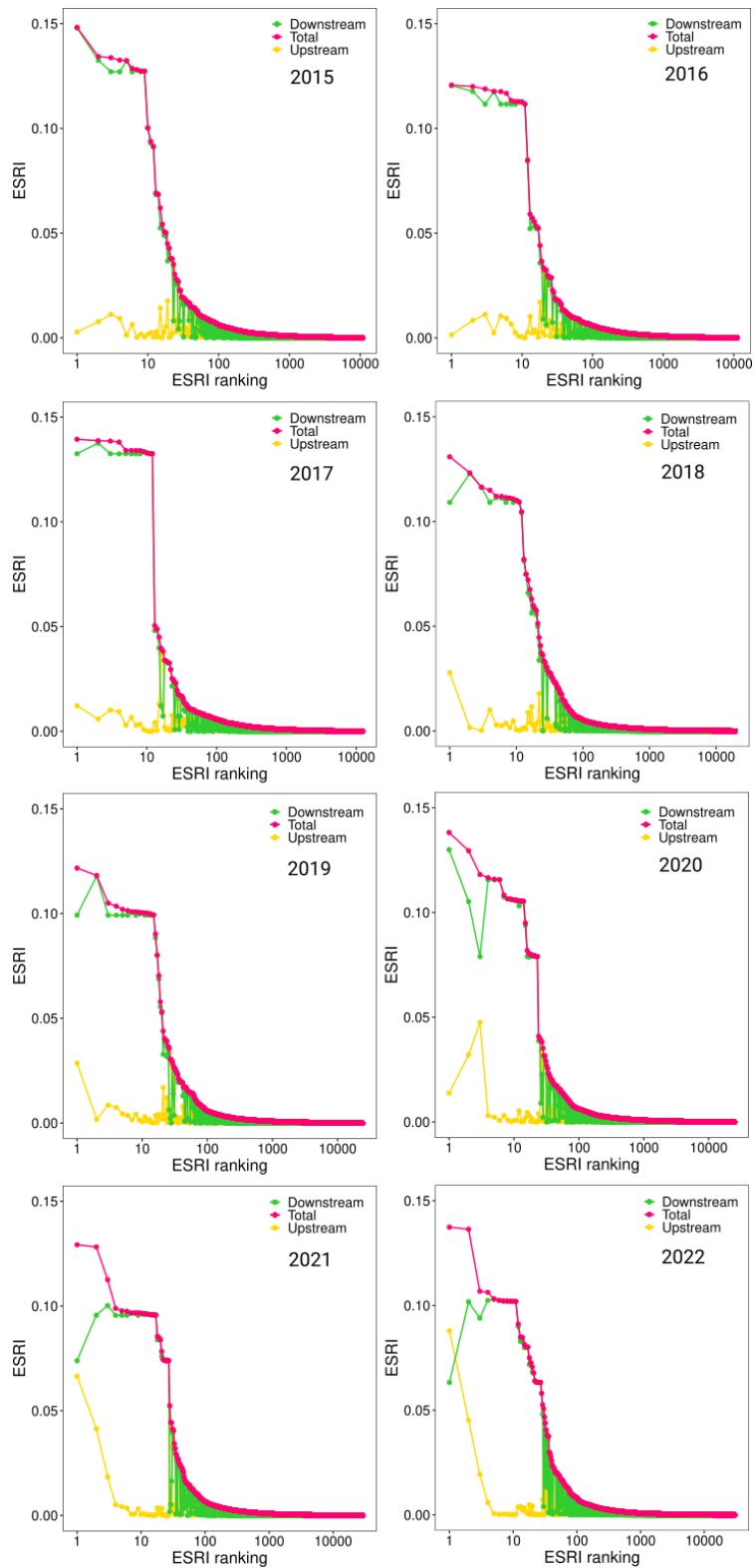

**Figure S10.** ESRI ranking for all empirical networks, from 2015 to 2022. In pink the total ESRI, in green the downstream values and in yellow the upstream values.

## S7 Degree and strength statistics for empirical and null model networks

For each year, below we show the following plots. The first row is the scatter plot for the empirical (blue) in-/out-degree vs in-/out-strength and the null model (in red) in-/out-degree vs in-/out-strength, obtained as averages over the s-GM ensemble. The plots also verify the fitness ansatz, according to which the strengths are proportional to the degrees and can be used to infer them. Panels in the second row represent the complementary cumulative distributions for the in- and out-degrees for both real (blue solid line) and null model (red dashed line) networks. The overlap between the curves highlights the optimal result of the null models in reproducing the empirical degrees. Third row panels instead represent the complementary cumulative distributions for the in- and out-strengths for both real (blue solid line) and null model (red dashed line) networks. The two curves overlap completely, as strength values are constrained by the s-GM.

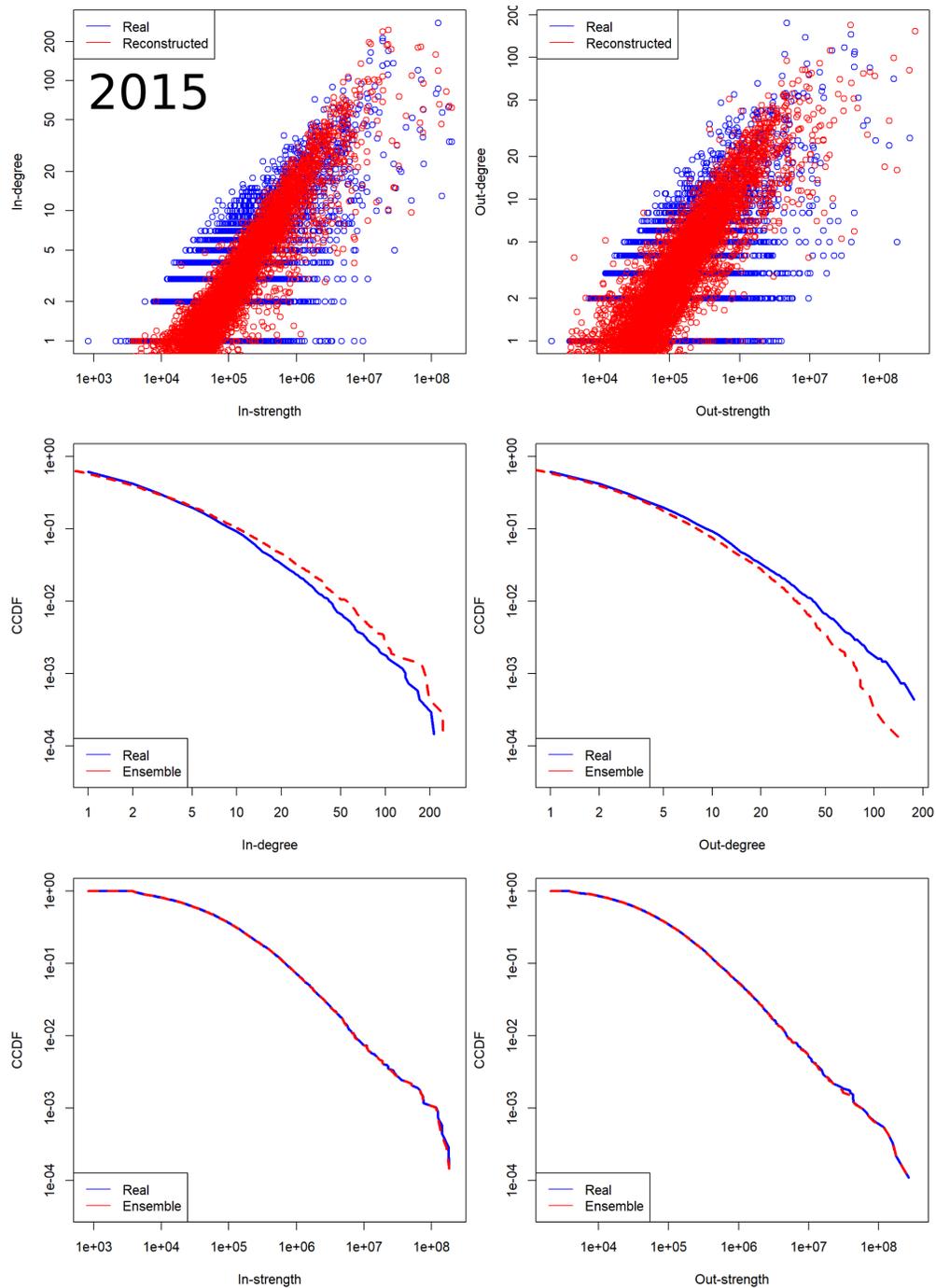

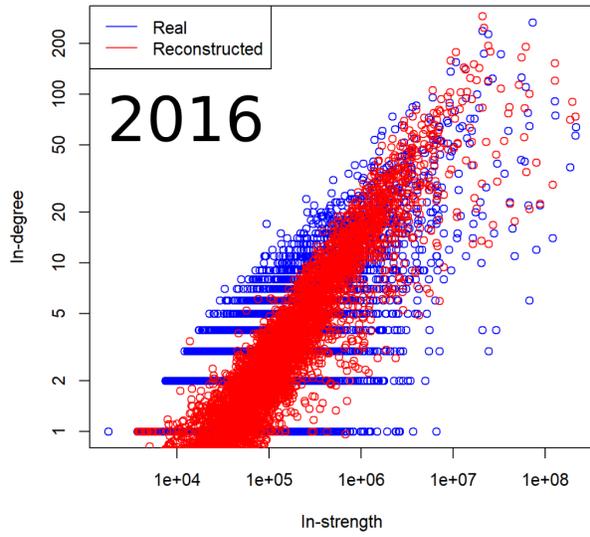
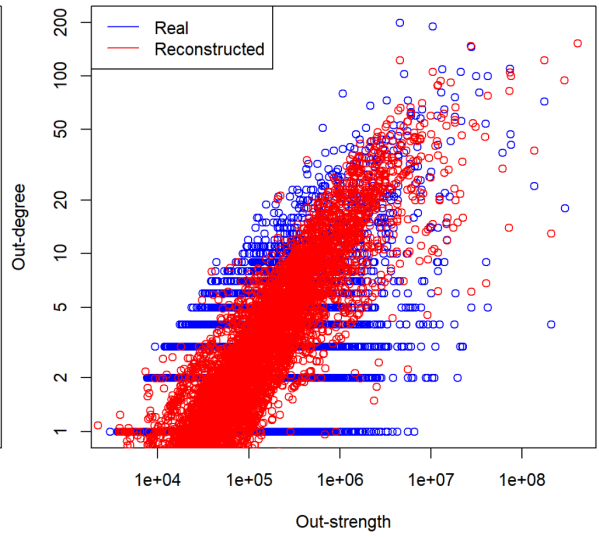
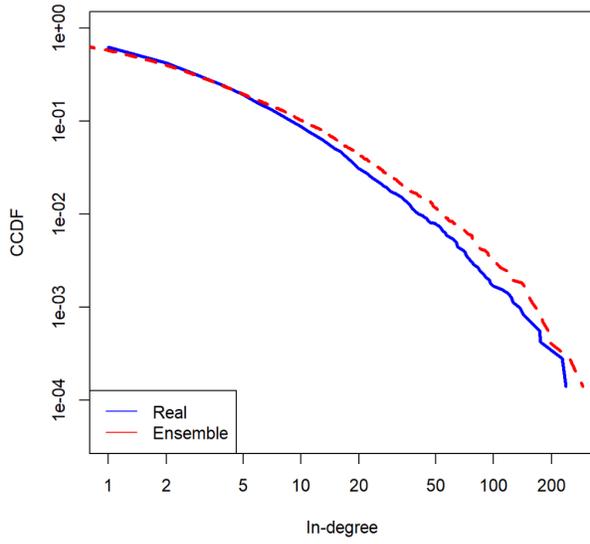
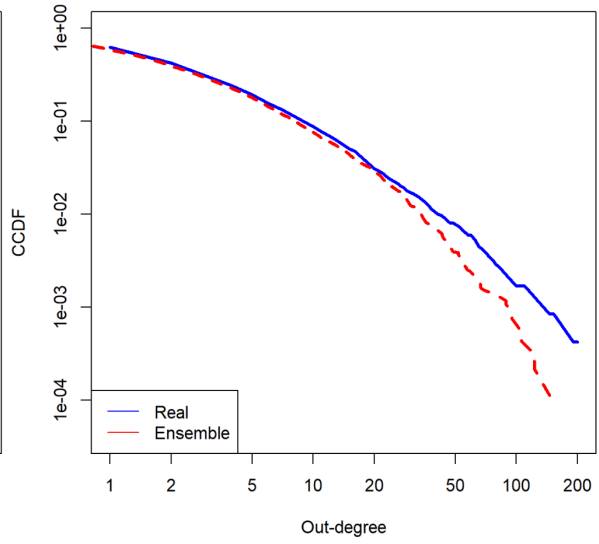
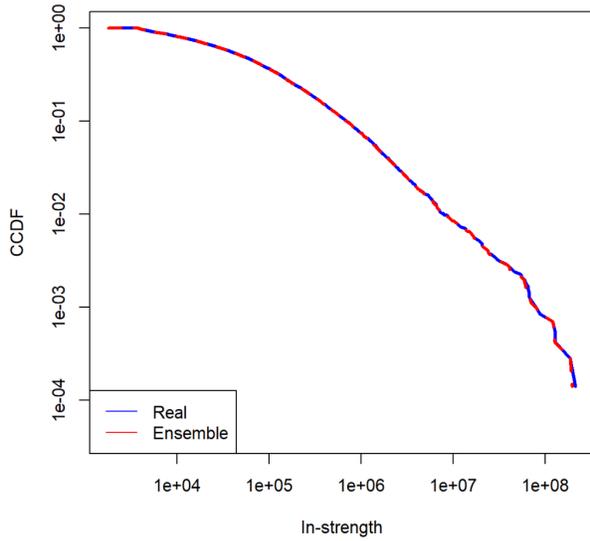
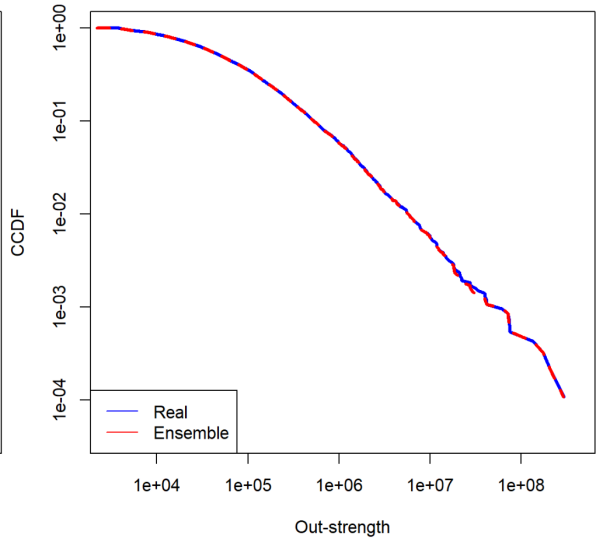

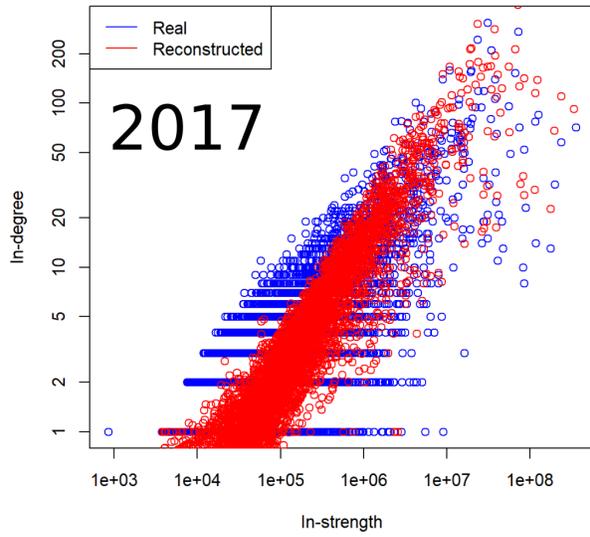
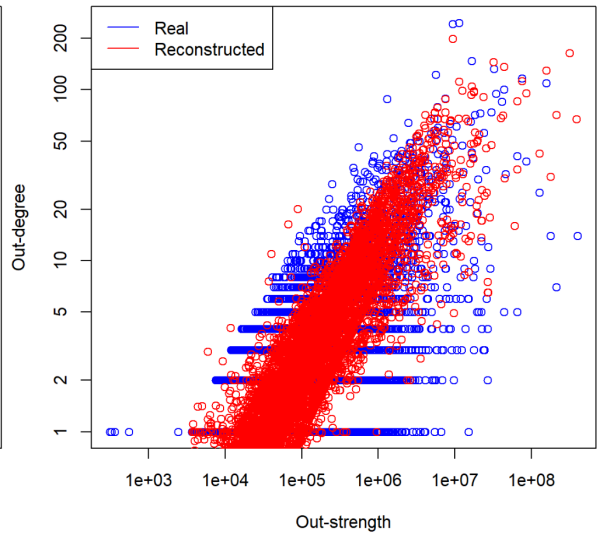
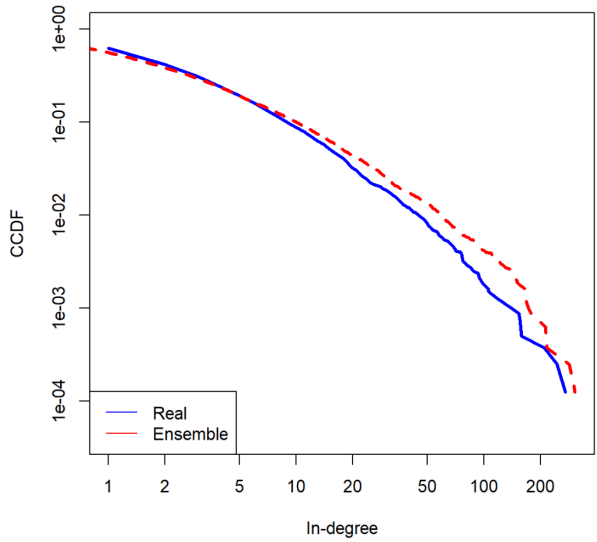
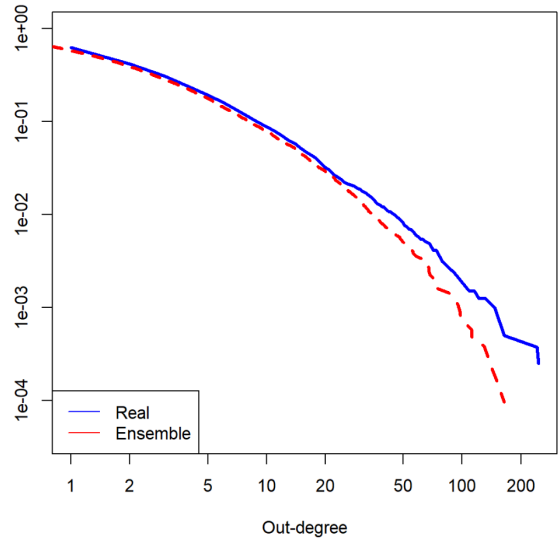
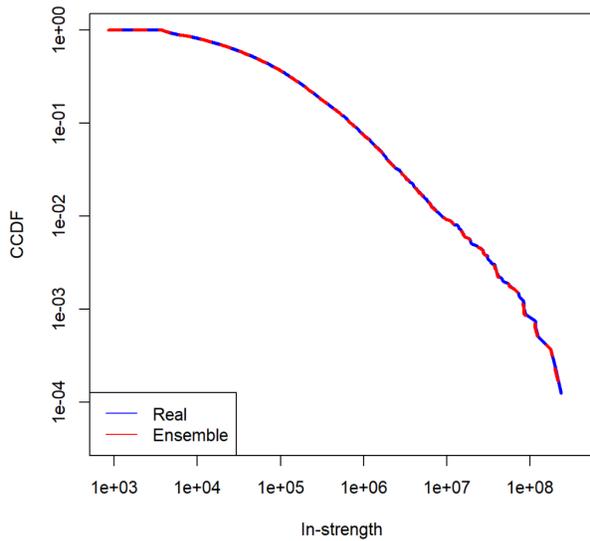
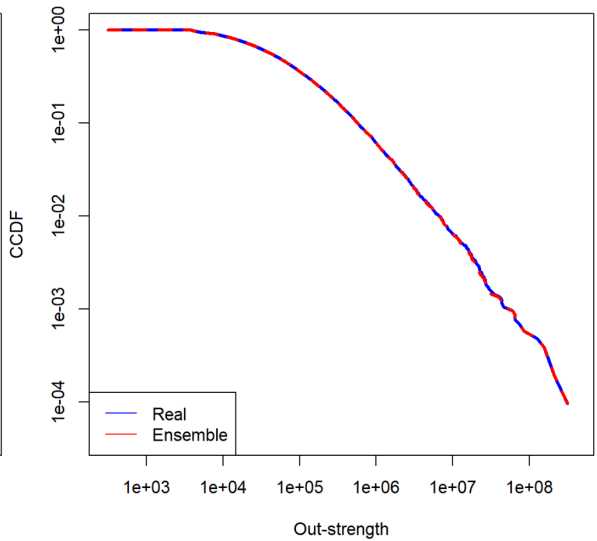

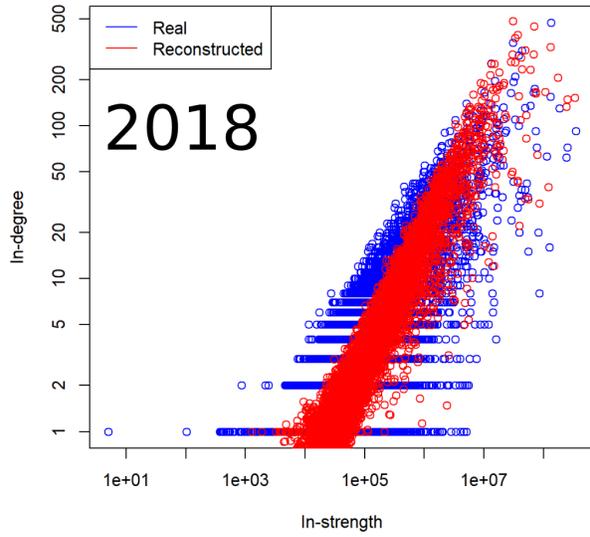
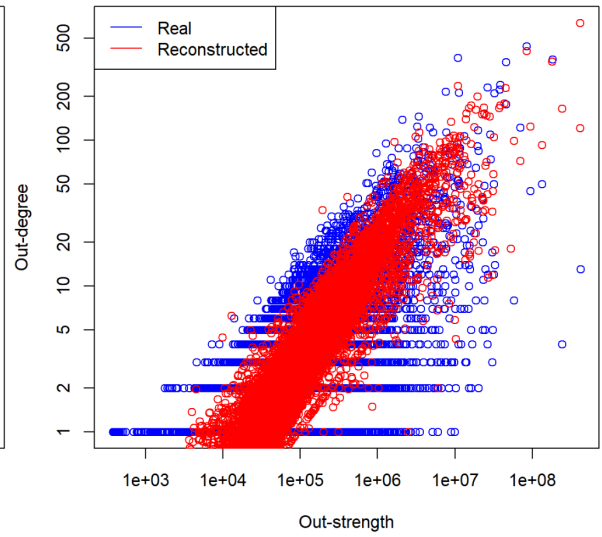
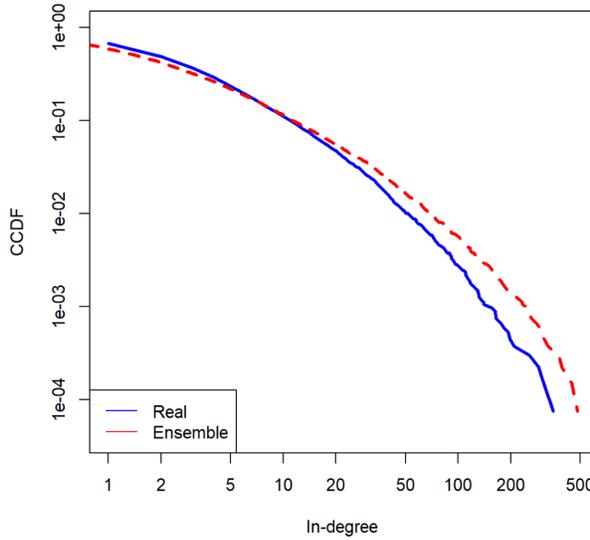
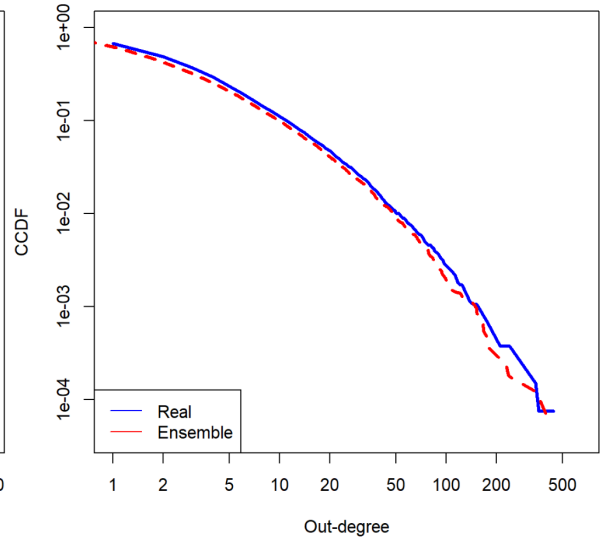
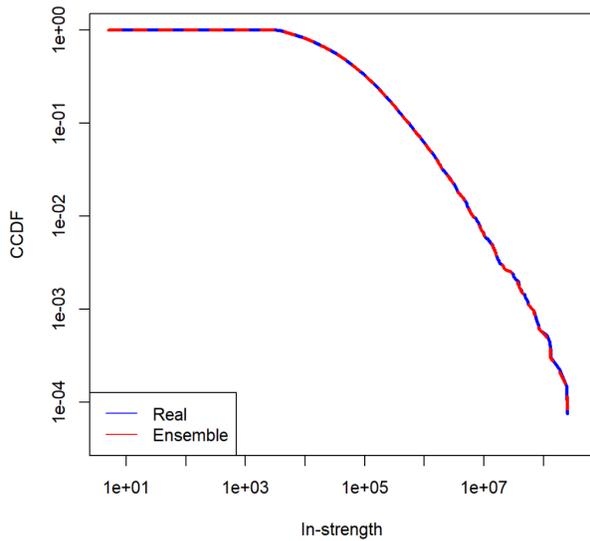
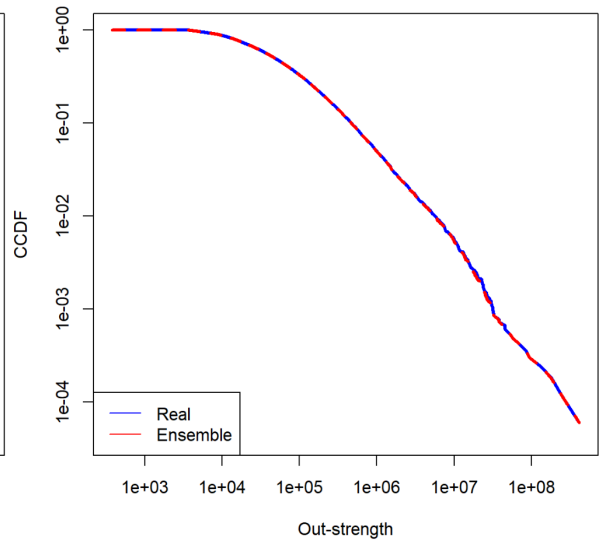

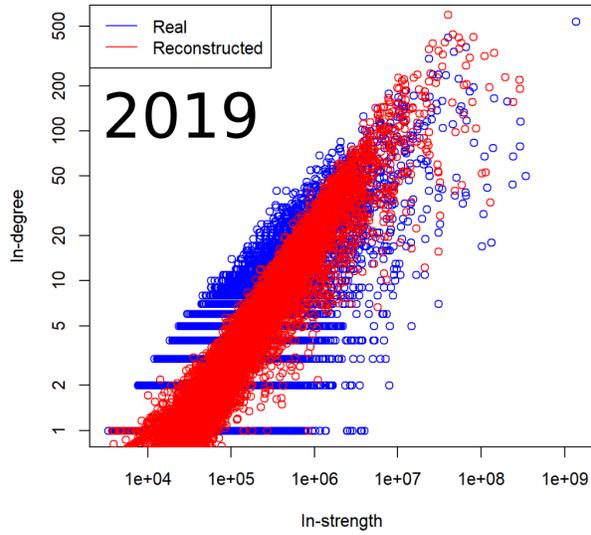
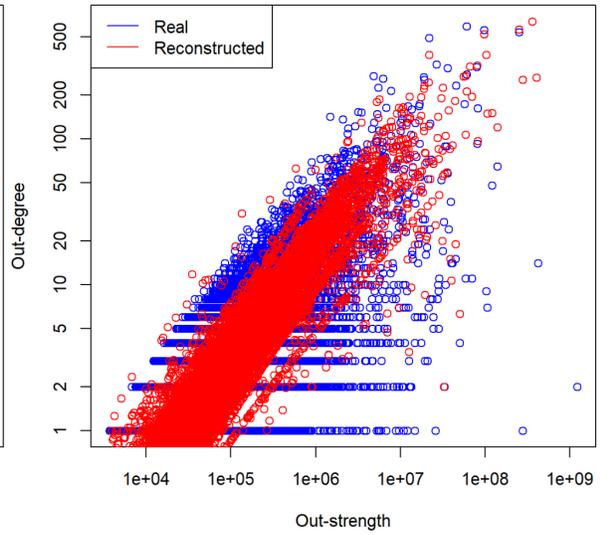
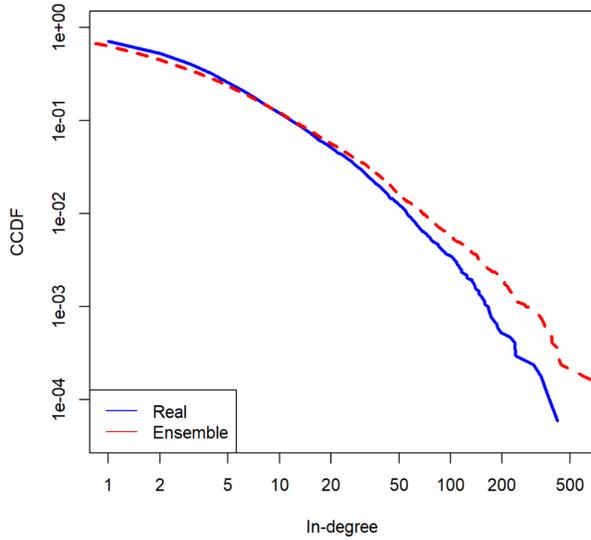
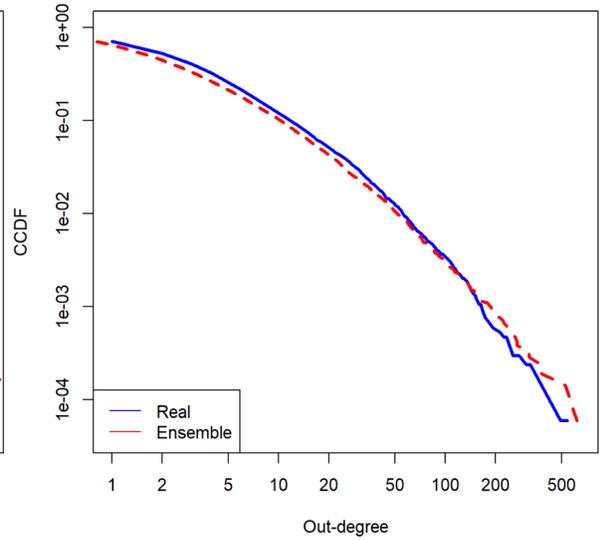
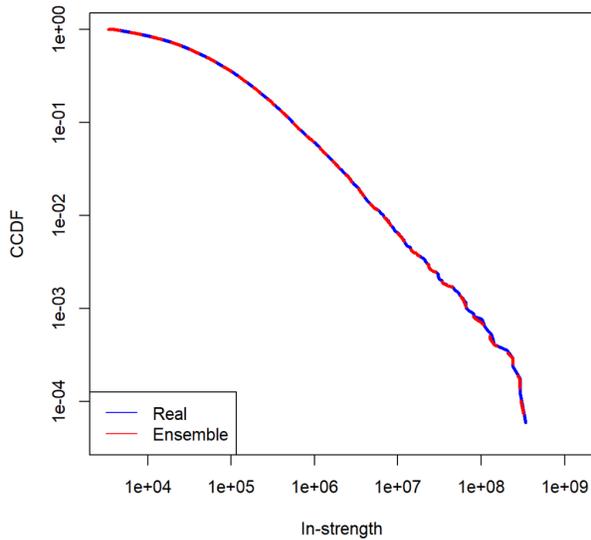
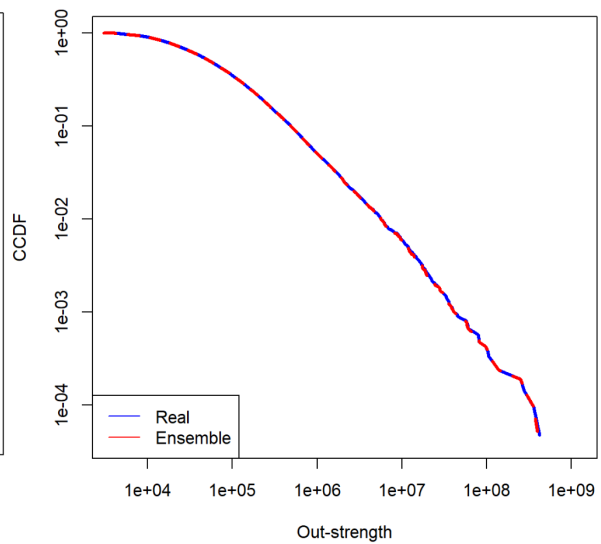

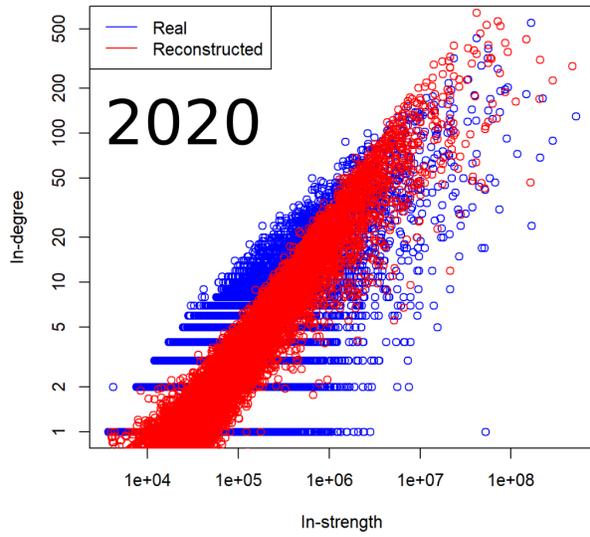
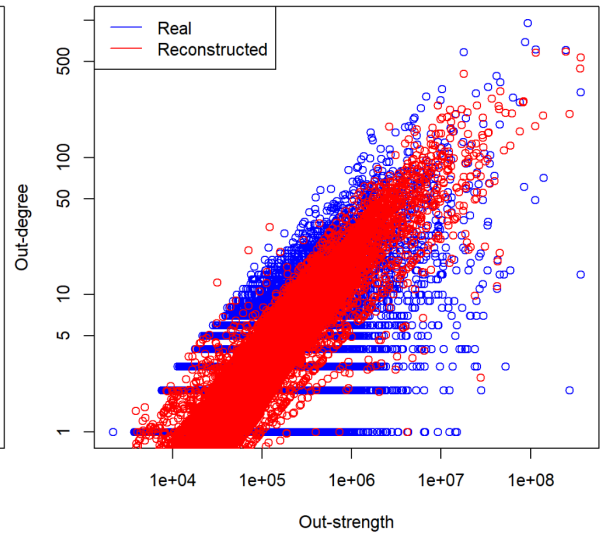
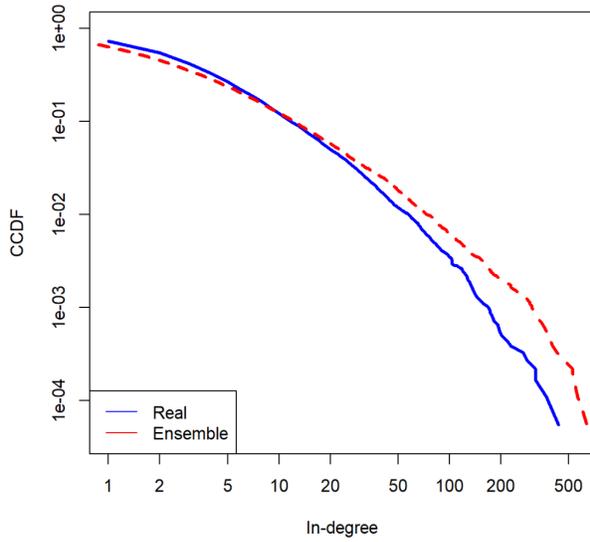
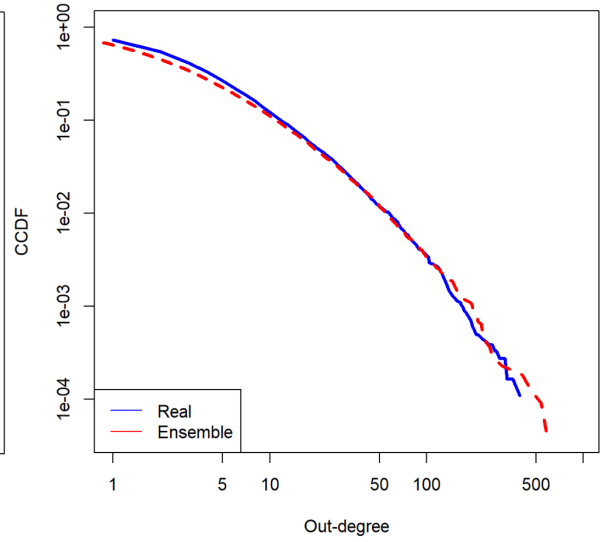
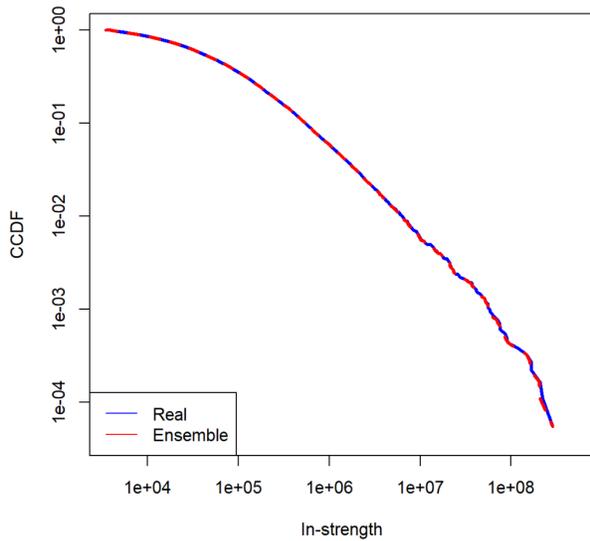
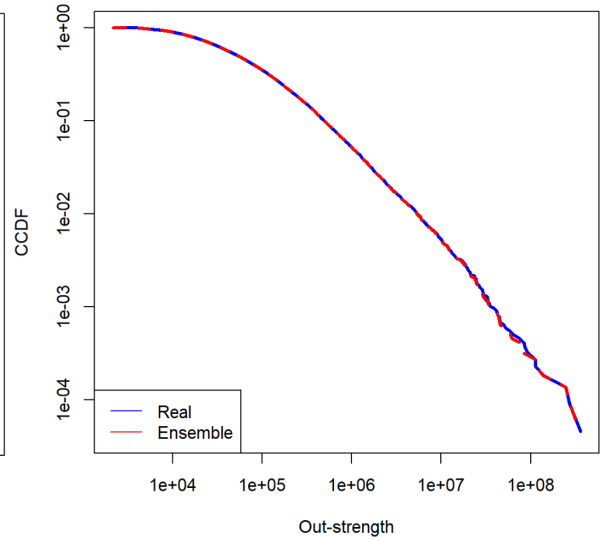

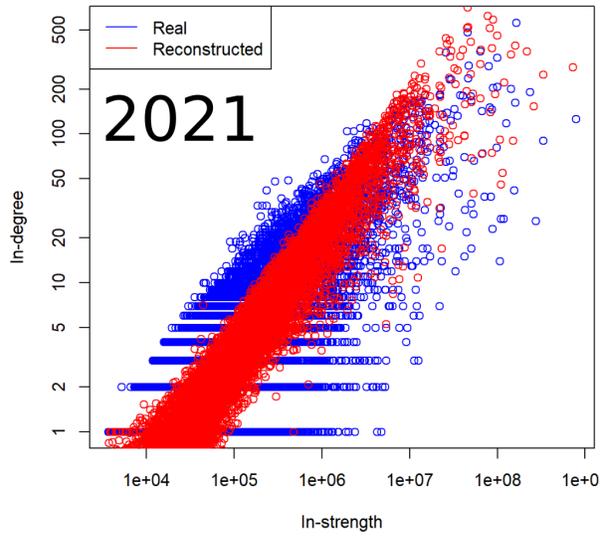
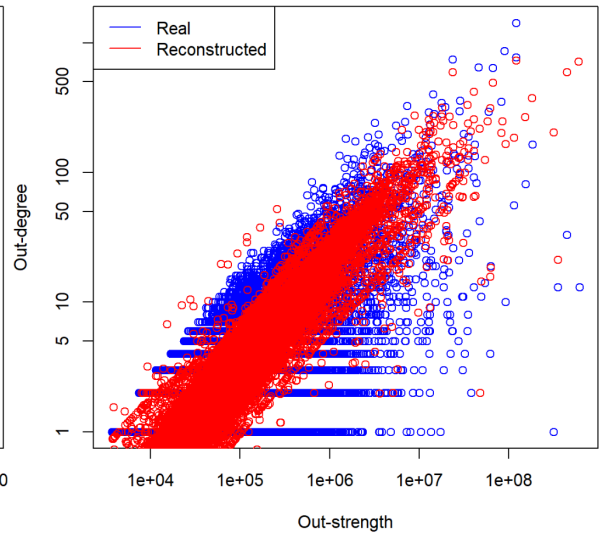
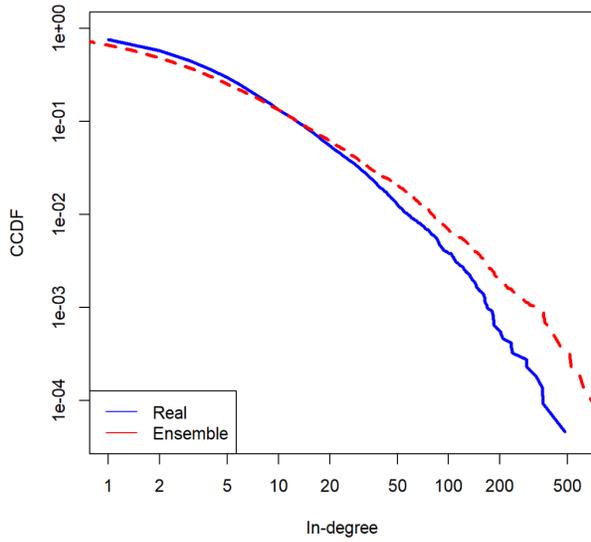
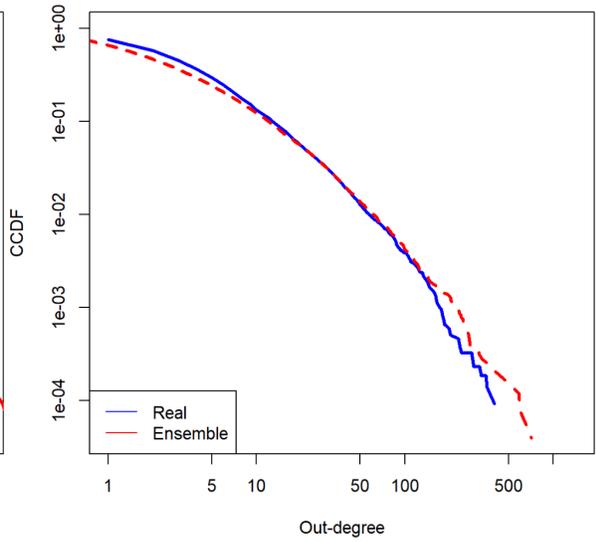
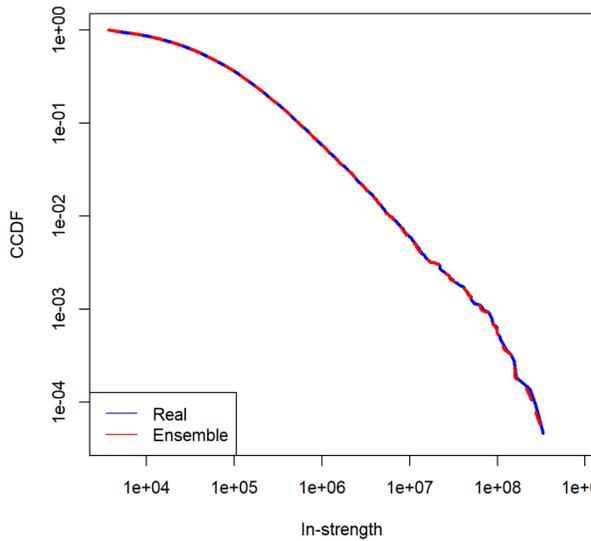
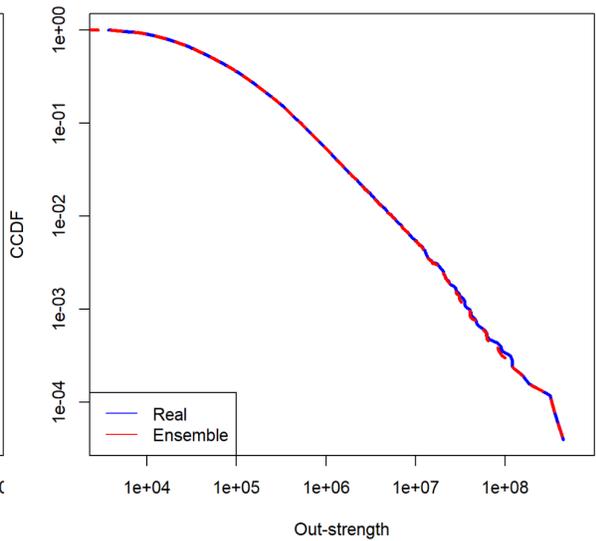

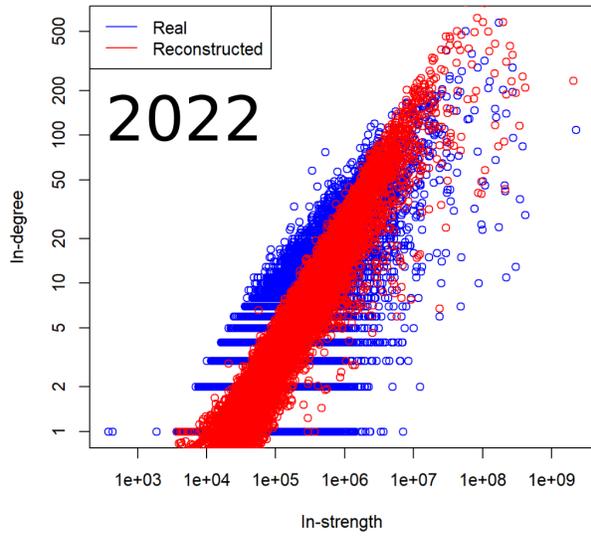
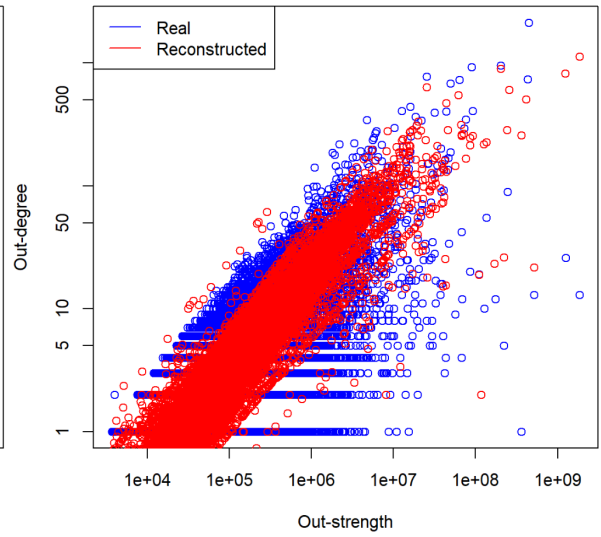
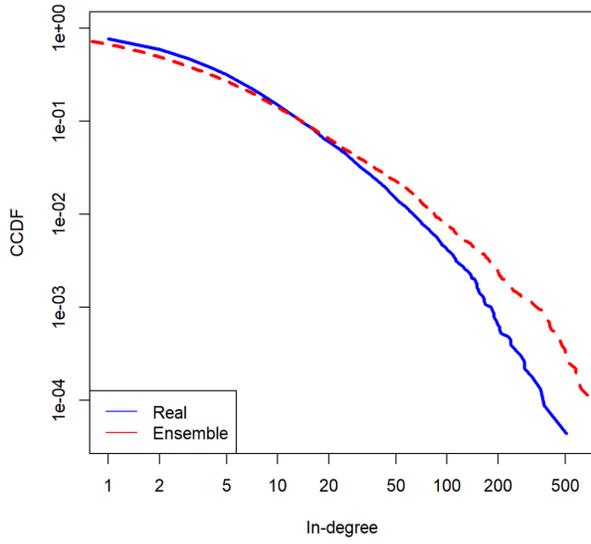
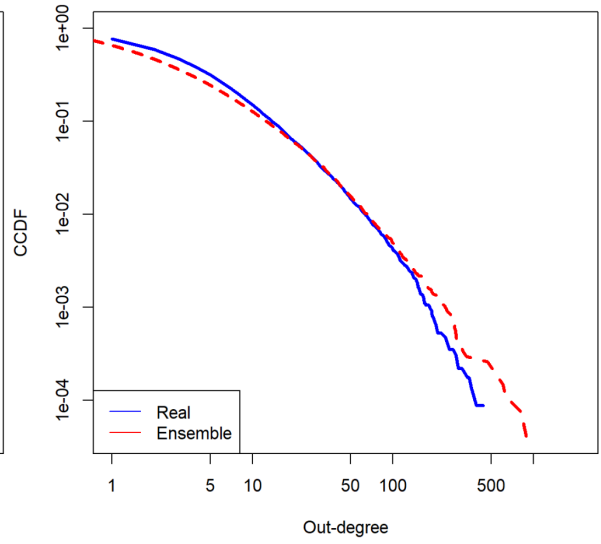
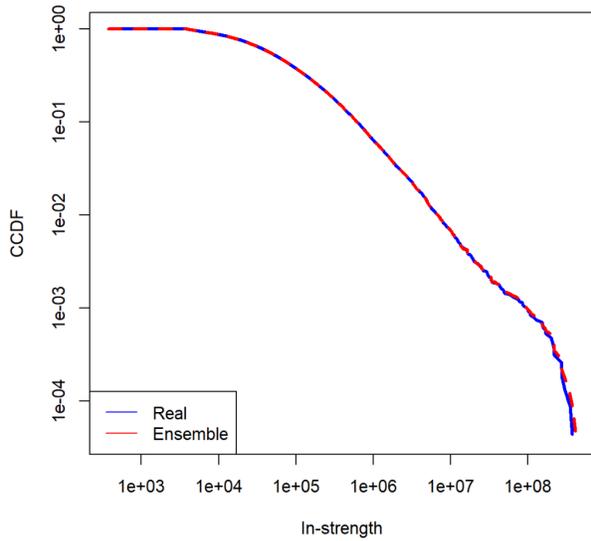
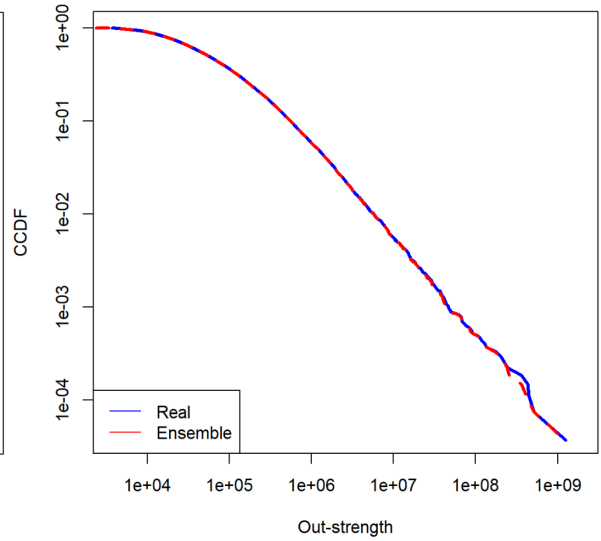

## S8 Empirical vs null model values of ESRI for 2015-2022

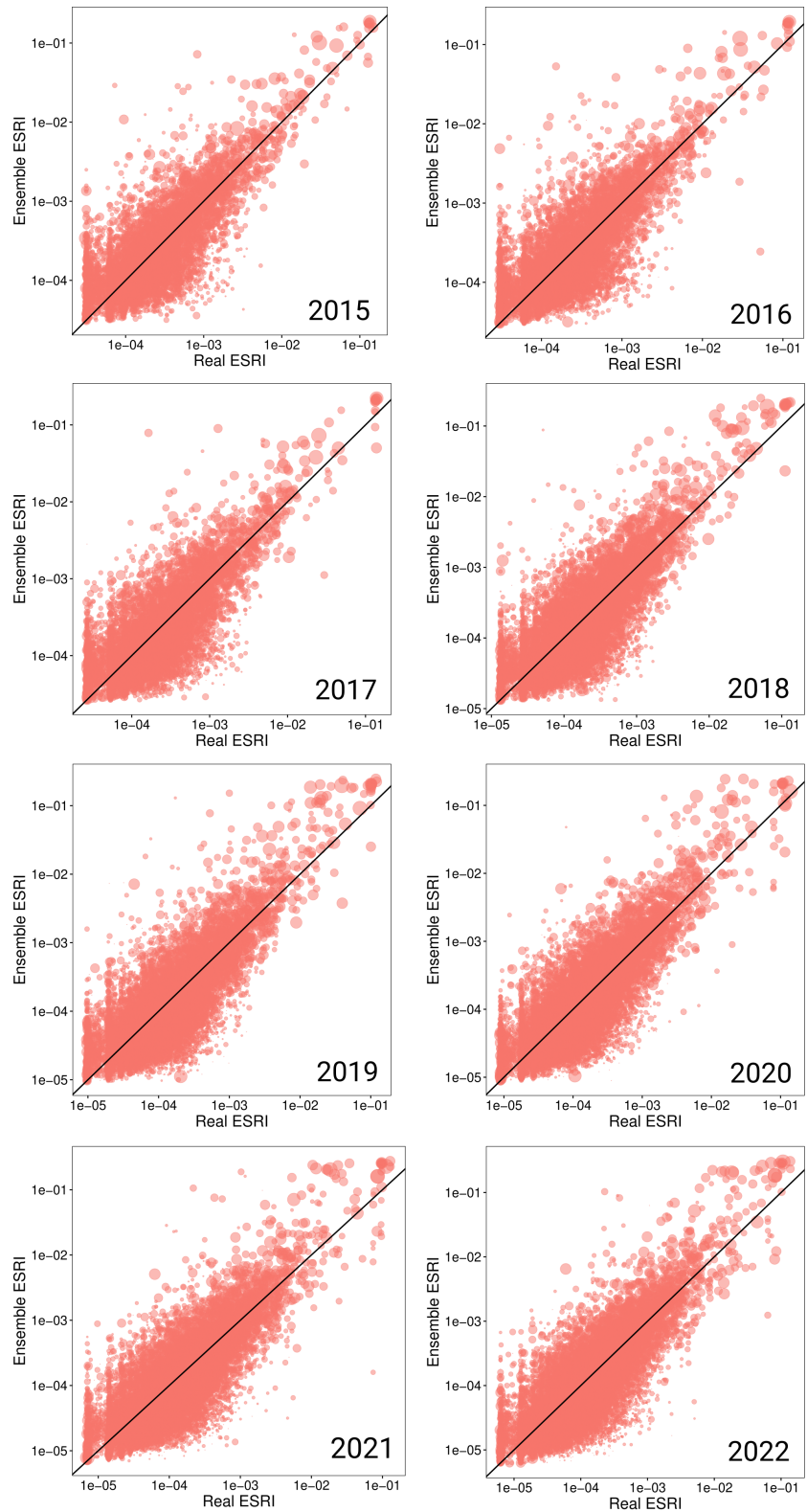

**Figure S8.** Scatter plots of empirical vs null model values of firm-level ESRI for all years, from 2015 to 2022. Dot size is proportional to the firm's number of employees.

## S9 Empirical vs null model rankings of ESRI for 2015-2022

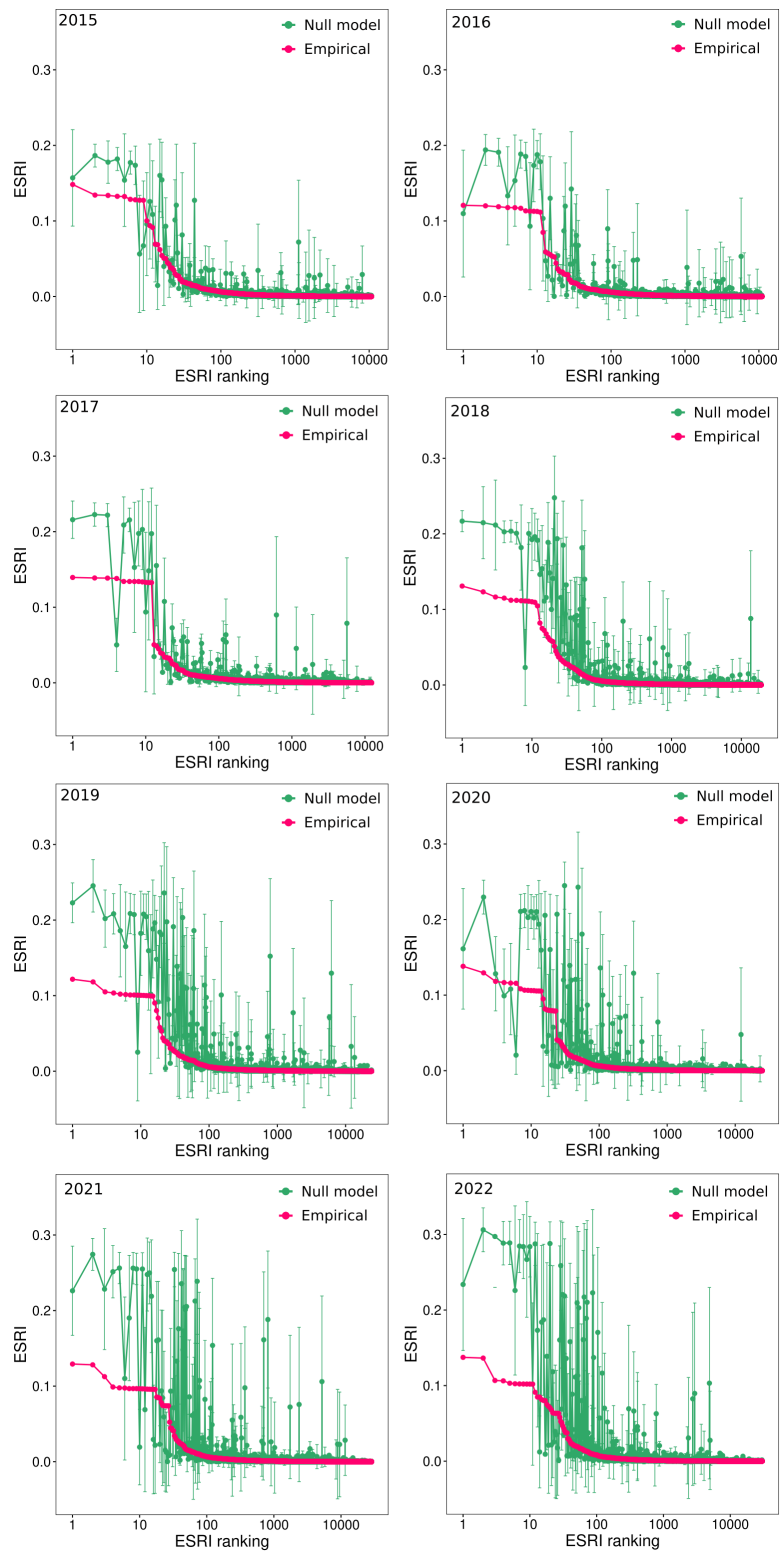

**Figure S9.** Ranking plots of empirical (pink) vs null model (green) values of ESRI for all years, from 2015 to 2022. As the years progress the two rankings become progressively divergent, underlying that the null model becomes less able to reproduce the empirical dynamics of firms.

## S10 Difference of ESRI rankings for empirical and null model networks

To quantify the mismatch between real and null model ESRI values shown in the rankings of the previous section, we compute for each firm the absolute value of the difference between empirical $ESRI$ and null model $\langle ESRI \rangle$ scores, and rank them for each year in decreasing order. The higher the curve, the greater the mismatch between the values. As can be seen by the plot below, the highest differences occur during the COVID years.

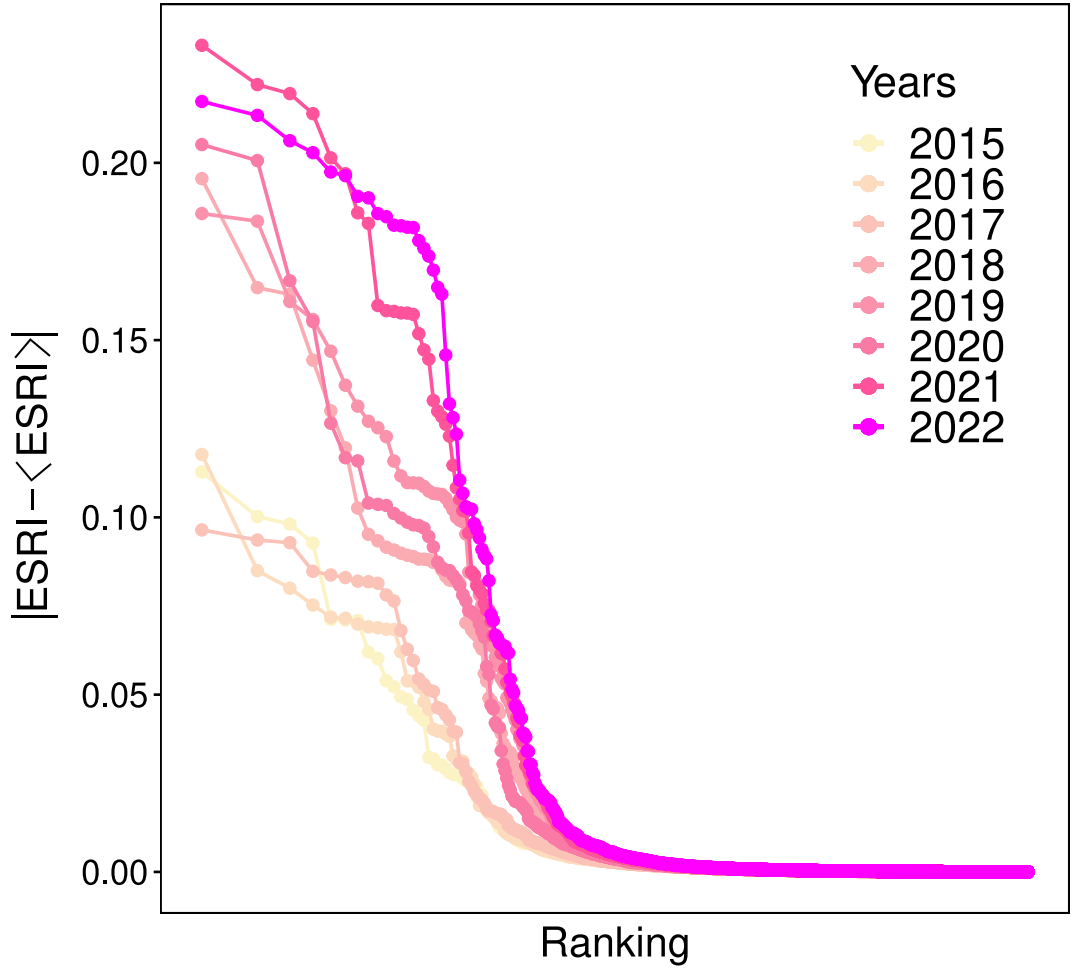

**Figure S10.** Absolute value of the difference between real and null model ESRI values for all firms, ranked in decreasing order for each year.

# S11 Distributions of ESRI values

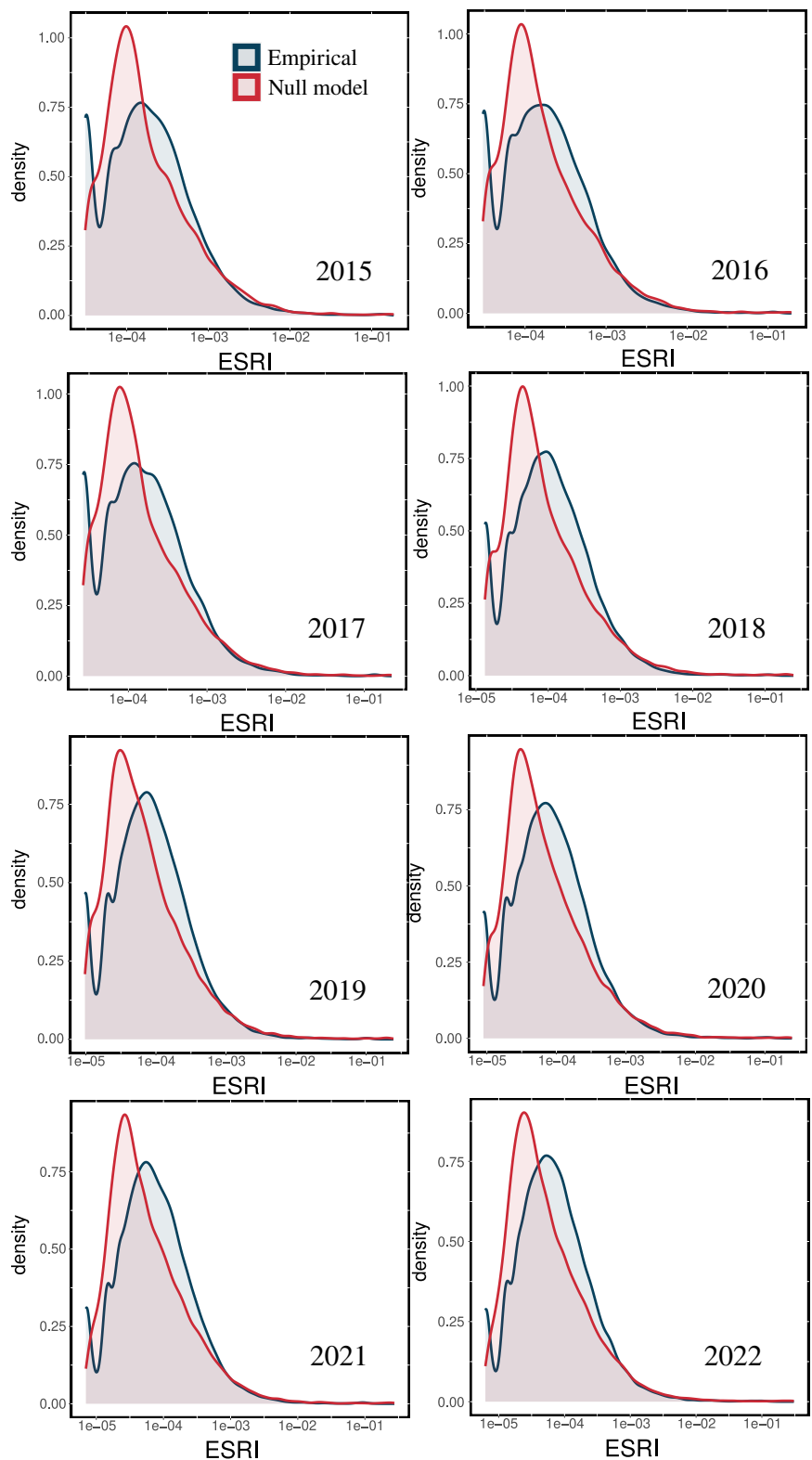

**Figure S11.** Distributions of ESRI values for all years, for empirical networks (blue) and null models (red).

## S12 Sectors of plateaux firms

| NACE4 code | Description |
| --- | --- |
| 0812 | Gravel, sand and clay mining |
| 2444 | Manufacture of copper |
| 2572 | Manufacture of locks and hinges |
| 2591 | Manufacture of steel drums and similar containers |
| 2593 | Manufacture of wire products |
| 2612 | Manufacture of loaded electronic boards |
| 2711 | Manufacture of electric motors, generators and transformers |
| 2712 | Manufacture of electricity distribution and control apparatus |
| 2732 | Manufacture of other electronic and electric wires and cables |
| 2733 | Manufacture of assemblies |
| 2740 | Manufacture of electric lighting equipment |
| 2751 | Manufacture of electric domestic appliances |
| 2790 | Manufacture of other electrical equipment |
| 2811 | Manufacture of engines and turbines (except aircraft and cycle engines) |
| 2812 | Manufacture of hydraulic and pneumatic equipment |
| 2815 | Manufacture of bearings, gears, gearing and driving elements |
| 2830 | Manufacture of agricultural and forestry machinery |
| 2849 | Manufacture of other machine tools |
| 2891 | Manufacture of machinery for metallurgy |
| 2895 | Manufacture of machinery for paper and paperboard production |
| 3317 | Repair and maintenance of other transport equipment |
| 3511 | Electricity generation |
| 3512 | Electricity transmission |
| 3513 | Electricity distribution |
| 3514 | Electricity generation from solar energy |
| 3522 | Gas distribution |
| 3523 | Gas trading |
| 3530 | Steam supply, air conditioning |
| 4644 | Wholesale of china and glassware and cleaning materials |
| 4671 | Wholesale of solid, liquid, and gaseous fuels and related products |
| 4672 | Wholesale of metals and metal ores |
| 4931 | Urban and suburban passenger land transport |
| 4910 | Passenger rail transport, interurban |
| 5310 | Postal activity |

**Table S12.** The 33 NACE4 sectors (codes and descriptions) found in the plateaux of ESRI ranking throughout the years.

## S13 Evolution of ESRI plateaux

Figure S13 A displays the ESRI evolution of the top 0.1% of most risky firms that appear in the plateaux of empirical and null model networks more than two years, averaged by NACE2 sectors. The color of each line represents the sector, and the corresponding description can be found in Figure S13 B. Figure S13 C displays the difference of ESRI when a firm last appeared in the plateau compared to when it first appeared. The consistently negative values for the empirical networks imply a decreasing trend, except for the the postal sector, and, to a lesser extent, the manufacture of computers and electronics. These two sectors in fact experienced a growth of their ESRI values during the years, underlining the relevance they gained over time and during the period of crisis and lockdown measures. We can hypothesize that an increasing number of firms (or individuals) bought computers needed for remote work, and incremented the use of courier services to ship and receive items. Instead, all null model sectors, except the manufacture of computers and electronics, show positive variations, implying an increasing trend over time.

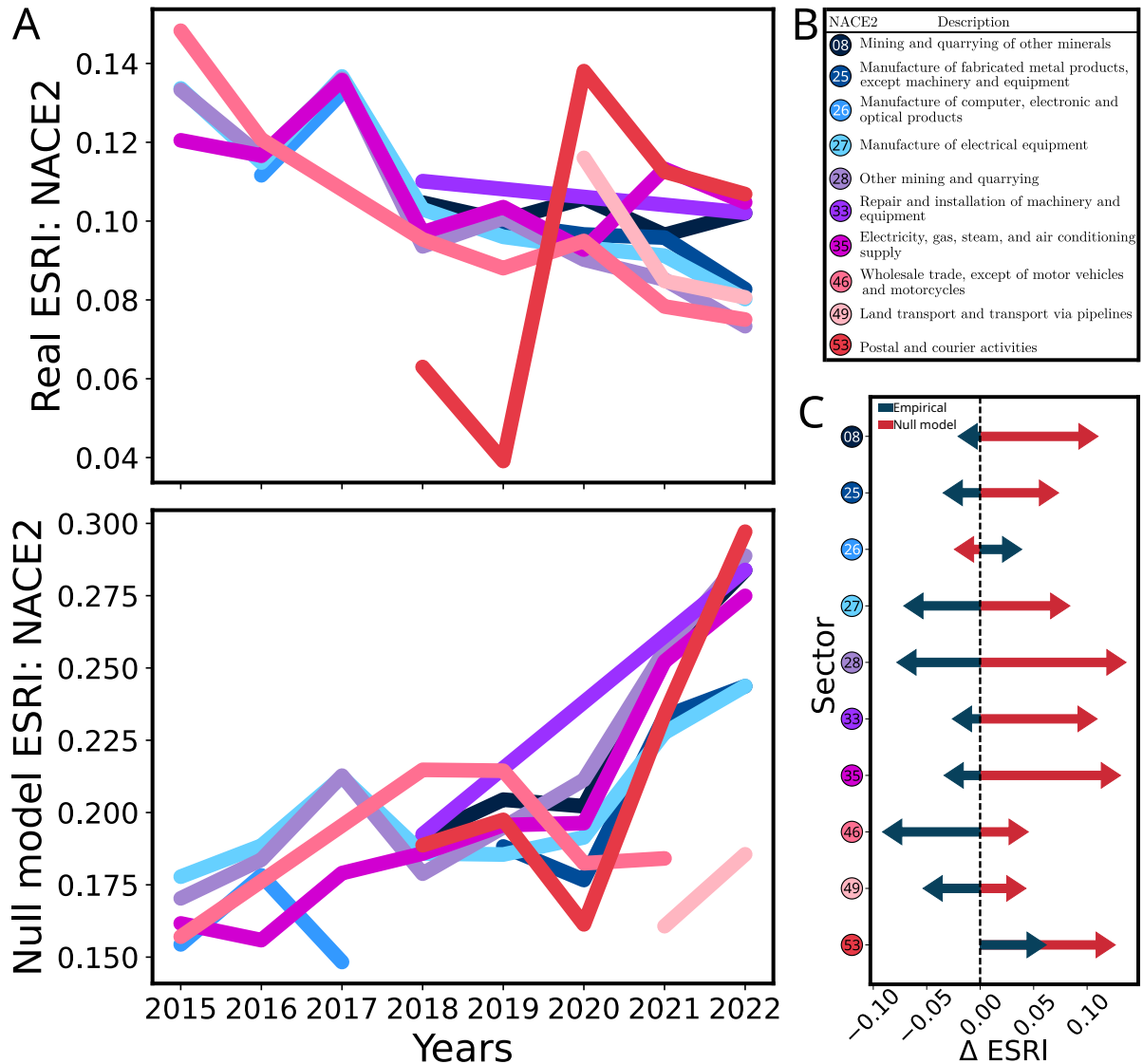

**Figure S13.** A) Yearly evolution of most risky firms belonging to the plateau for more than two years, present both in empirical and and null model networks, colored by their NACE2 code. Top panel represents empirical values, while bottom panel displays null model values. B) Color legend of NACE2 codes with description. C) Difference between ESRI values of a firm when it last appeared in the plateau and when it first appeared, blue arrows for empirical values and red arrows for null model.

## S14 Correlation of firm features

Table S13 summarizes the firm features employed in the regression analysis and what each feature represents. Figure S14 shows the Pearson correlation coefficient between each pair of features, for all years. During COVID years, import and export appear to be less correlated with all of the other variables, as can be expected given the import and export bans introduced to contain the pandemic.

| Feature | Description |
|---|---|
| $k_{in}$ | in-degree (number of suppliers) of the firm |
| $k_{out}$ | out-degree (number of customers) of the firm |
| $s_{in}$ | in-strength (total value of purchases) of the firm |
| $s_{out}$ | out-strength (total value of sales) of the firm |
| $emp$ | number of employees of the firm |
| $ess$ | essentiality score of the firm |
| $ESRI$ | empirical ESRI of the firm |
| $ESRI_{model}$ | null model ESRI of the firm |
| $import$ | value of the yearly import of the firm |
| $export$ | value of the yearly export of the firm |

**Table S13.** Description of the 10 features employed in the regression analysis.

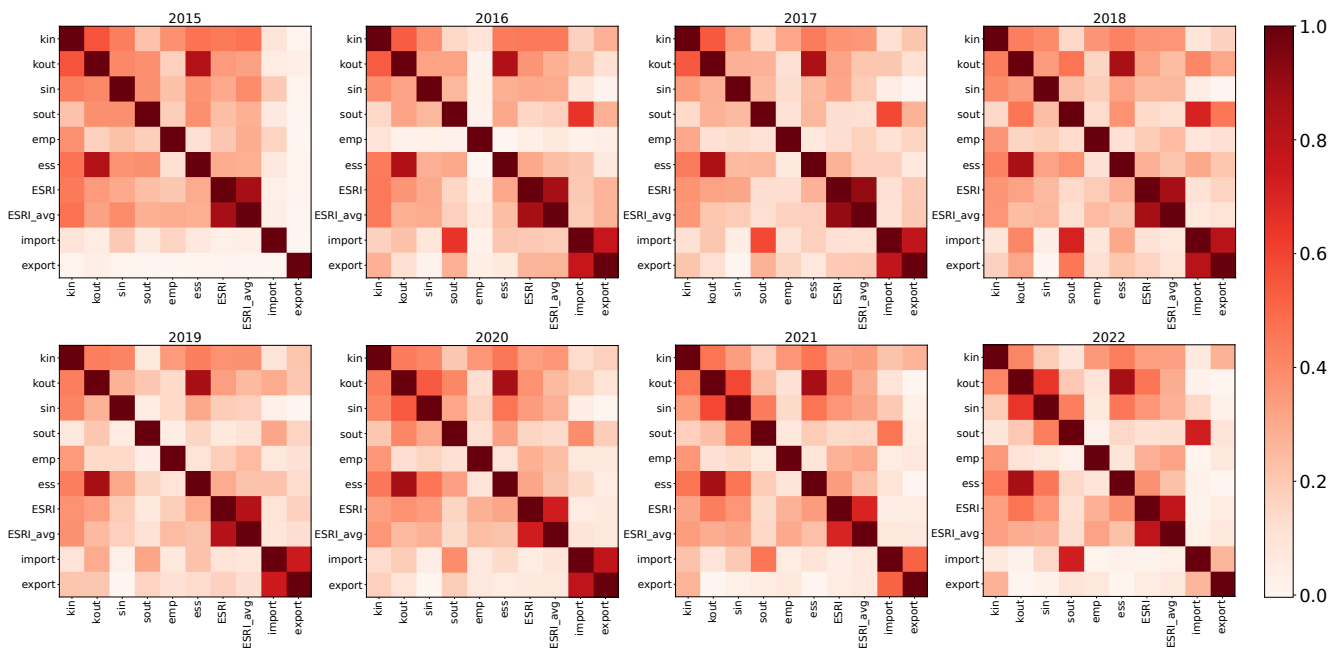

**Figure S14.** Person correlation coefficients between all pairs of firm variables, throughout the entire time span of data.

## S15 Essentiality score vs ESRI

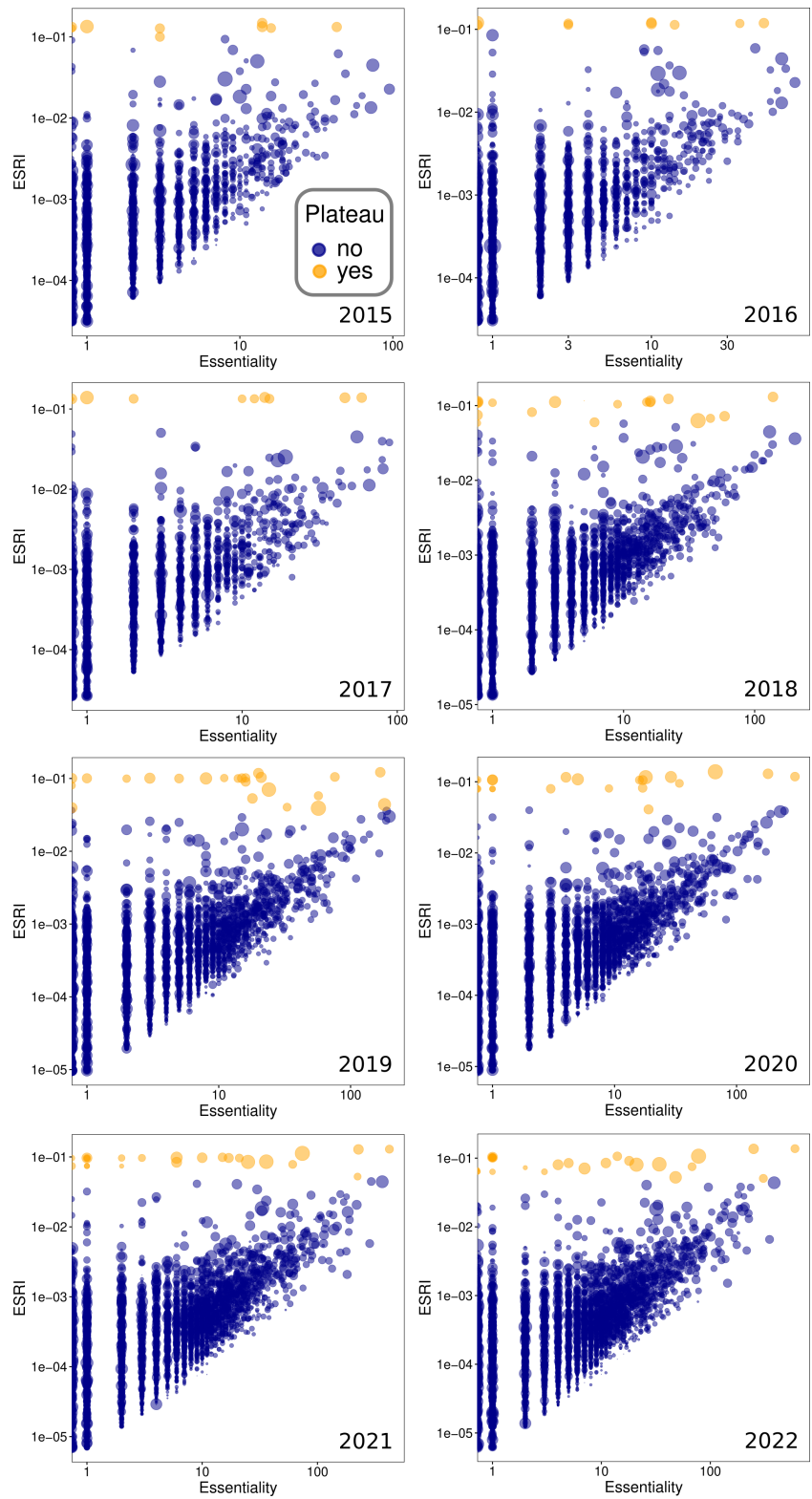

**Figure S15.** Scatter plot of essentiality score vs ESRI. Dot size is proportional to the firm's out-strength. Orange points are firms that belong to the ESRI plateau of that given year.

## S16 Pooled OLS regressions with year fixed effects

|  | ESRI | | | | |
|---|---|---|---|---|---|
|  | (1) | (2) | (3) | (4) | (5) |
| $ESRI_{model}$ | 0.733*** | 0.724*** | 0.704*** | 0.696*** | 0.684*** |
|  | (0.002) | (0.002) | (0.002) | (0.002) | (0.002) |
| import |  |  |  |  | 0.011*** |
|  |  |  |  |  | (0.000) |
| export |  |  |  |  | 0.004*** |
|  |  |  |  |  | (0.001) |
| emp |  |  |  | 0.027*** | 0.009*** |
|  |  |  |  | (0.002) | (0.002) |
| year = 2016 |  | 0.000 | −0.001 | −0.001 | −0.010** |
|  |  | (0.005) | (0.005) | (0.005) | (0.005) |
| year = 2017 |  | −0.002 | −0.004 | −0.004 | −0.014*** |
|  |  | (0.005) | (0.005) | (0.005) | (0.005) |
| year = 2018 |  | −0.027*** | −0.025*** | −0.025*** | −0.034*** |
|  |  | (0.004) | (0.004) | (0.004) | (0.004) |
| year = 2019 |  | −0.041*** | −0.040*** | −0.038*** | −0.050*** |
|  |  | (0.004) | (0.004) | (0.004) | (0.004) |
| year = 2020 |  | −0.043*** | −0.043*** | −0.041*** | −0.054*** |
|  |  | (0.004) | (0.004) | (0.004) | (0.004) |
| year = 2021 |  | −0.047*** | −0.045*** | −0.043*** | −0.057*** |
|  |  | (0.004) | (0.004) | (0.004) | (0.004) |
| year = 2022 |  | −0.053*** | −0.050*** | −0.047*** | −0.063*** |
|  |  | (0.004) | (0.004) | (0.004) | (0.004) |
| Constant | −1.018*** | −1.023*** | −1.276*** | −1.338*** | −1.364*** |
|  | (0.006) | (0.007) | (0.043) | (0.043) | (0.043) |
| Observations | 160,895 | 160,895 | 160,895 | 157,612 | 157,612 |
| $R^2$ | 0.587 | 0.588 | 0.614 | 0.618 | 0.621 |
| Adjusted $R^2$ | 0.587 | 0.588 | 0.613 | 0.617 | 0.620 |
| Residual Std. Error | 0.364 (df = 160893) | 0.363 (df = 160886) | 0.352 (df = 160325) | 0.351 (df = 157041) | 0.350 (df = 157039) |
| F Statistic | 228,593.900*** (df = 1; 160893) | 28,694.400*** (df = 8; 160886) | 448.369*** (df = 569; 160325) | 446.290*** (df = 570; 157041) | 450.000*** (df = 572; 157039) |

**Table S16.** Pooled OLS regressions with year fixed effects. *** Significant at the 1% level. ** Significant at the 5% level. * Significant at the 10% level.

## S17 Cross-sectional regressions for import

| | ESRI | | | | | | | |
|---|---|---|---|---|---|---|---|---|
| | (1) | (2) | (3) | (4) | (5) | (6) | (7) | (8) |
| $ESRI_{model}$ | 0.425*** | 0.399*** | 0.371*** | 0.379*** | 0.373*** | 0.389*** | 0.372*** | 0.350*** |
| | (0.016) | (0.016) | (0.015) | (0.011) | (0.011) | (0.011) | (0.010) | (0.010) |
| $import$ | −0.003 | 0.010 | 0.010 | 0.021*** | 0.027*** | 0.034*** | 0.034*** | 0.020*** |
| | (0.009) | (0.007) | (0.007) | (0.006) | (0.006) | (0.006) | (0.005) | (0.005) |
| $s_{out}$ | −0.017 | −0.001 | 0.022** | 0.033*** | 0.048*** | 0.043*** | 0.052*** | 0.054*** |
| | (0.011) | (0.011) | (0.010) | (0.008) | (0.008) | (0.008) | (0.008) | (0.008) |
| $s_{in}$ | 0.049*** | 0.045*** | 0.045*** | 0.038*** | 0.038*** | 0.040*** | 0.039*** | 0.045*** |
| | (0.002) | (0.002) | (0.002) | (0.001) | (0.001) | (0.001) | (0.001) | (0.001) |
| $ess$ | 0.529*** | 0.522*** | 0.513*** | 0.488*** | 0.475*** | 0.472*** | 0.473*** | 0.479*** |
| | (0.017) | (0.019) | (0.017) | (0.012) | (0.011) | (0.011) | (0.010) | (0.010) |
| $emp$ | −0.006 | −0.022*** | −0.017*** | −0.007 | 0.012** | 0.003 | 0.005 | 0.015*** |
| | (0.007) | (0.007) | (0.006) | (0.005) | (0.005) | (0.005) | (0.005) | (0.005) |
| $plateau$ | 1.200*** | 1.183*** | 1.082*** | 1.083*** | 0.897*** | 1.143*** | 1.160*** | 1.191*** |
| | (0.135) | (0.131) | (0.138) | (0.090) | (0.080) | (0.089) | (0.092) | (0.089) |
| $import : s_{out}$ | 0.000 | −0.000 | −0.001 | −0.006*** | −0.006*** | −0.008*** | −0.009*** | −0.007*** |
| | (0.002) | (0.002) | (0.002) | (0.001) | (0.001) | (0.001) | (0.001) | (0.001) |
| $import : s_{in}$ | 0.001 | 0.001* | 0.002*** | 0.003*** | 0.002*** | 0.002*** | 0.004*** | 0.004*** |
| | (0.001) | (0.000) | (0.000) | (0.000) | (0.000) | (0.000) | (0.000) | (0.000) |
| $import : ess$ | 0.006 | −0.002 | −0.004 | 0.002 | 0.002 | 0.006*** | 0.004** | 0.003 |
| | (0.004) | (0.004) | (0.004) | (0.003) | (0.002) | (0.002) | (0.002) | (0.002) |
| $Constant$ | −2.714*** | −2.655*** | −2.785*** | −2.858*** | −3.218*** | −2.976*** | −3.208*** | −3.259*** |
| | (0.176) | (0.252) | (0.210) | (0.180) | (0.132) | (0.129) | (0.134) | (0.125) |
| Observations | 9,000 | 9,253 | 10,308 | 16,474 | 20,604 | 21,579 | 24,909 | 25,829 |
| $R^2$ | 0.630 | 0.613 | 0.623 | 0.622 | 0.631 | 0.630 | 0.630 | 0.642 |
| Adjusted $R^2$ | 0.609 | 0.592 | 0.605 | 0.610 | 0.621 | 0.621 | 0.622 | 0.634 |
| Residual Std. Error | 0.318 (df = 8522) | 0.322 (df = 8783) | 0.318 (df = 9835) | 0.320 (df = 15963) | 0.317 (df = 20087) | 0.321 (df = 21060) | 0.328 (df = 24378) | 0.329 (df = 25299) |
| F Statistic | 30.375*** (df = 477; 8522) | 29.662*** (df = 469; 8783) | 34.504*** (df = 472; 9835) | 51.526*** (df = 510; 15963) | 66.464*** (df = 516; 20087) | 69.282*** (df = 518; 21060) | 78.276*** (df = 530; 24378) | 85.638*** (df = 529; 25299) |

**Table S17.** Cross sectional regressions for import. *** Significant at the 1% level. ** Significant at the 5% level. * Significant at the 10% level.

## S18 Cross-sectional regressions for export

| | ESRI | | | | | | | |
|---|---|---|---|---|---|---|---|---|
| | (1) | (2) | (3) | (4) | (5) | (6) | (7) | (8) |
| $ESRI_{model}$ | 0.427*** | 0.395*** | 0.369*** | 0.380*** | 0.371*** | 0.387*** | 0.371*** | 0.354*** |
| | (0.016) | (0.016) | (0.015) | (0.011) | (0.011) | (0.011) | (0.010) | (0.010) |
| $export$ | −0.031 | −0.016* | −0.008 | 0.003 | 0.007 | 0.012*** | 0.015** | 0.000 |
| | (0.021) | (0.009) | (0.009) | (0.008) | (0.008) | (0.008) | (0.008) | (0.007) |
| $s_{out}$ | −0.018* | 0.000 | 0.023** | 0.027*** | 0.045*** | 0.038*** | 0.046*** | 0.046*** |
| | (0.010) | (0.011) | (0.010) | (0.008) | (0.008) | (0.008) | (0.008) | (0.008) |
| $s_{in}$ | 0.050*** | 0.045*** | 0.046*** | 0.040*** | 0.039*** | 0.041*** | 0.041*** | 0.047*** |
| | (0.002) | (0.002) | (0.002) | (0.001) | (0.001) | (0.001) | (0.001) | (0.001) |
| $ess$ | 0.541*** | 0.534*** | 0.522*** | 0.498*** | 0.478*** | 0.481*** | 0.479*** | 0.483*** |
| | (0.016) | (0.017) | (0.016) | (0.012) | (0.010) | (0.010) | (0.009) | (0.009) |
| $emp$ | −0.004 | −0.011* | −0.008 | −0.003 | 0.017*** | 0.007 | 0.011** | 0.021*** |
| | (0.007) | (0.007) | (0.006) | (0.005) | (0.005) | (0.005) | (0.005) | (0.005) |
| $plateau$ | 1.176*** | 1.138*** | 1.078*** | 1.063*** | 0.853*** | 1.140*** | 1.153*** | 1.179*** |
| | (0.135) | (0.132) | (0.138) | (0.090) | (0.081) | (0.089) | (0.092) | (0.089) |
| $export:s_{out}$ | 0.006 | 0.003 | 0.002 | −0.002 | −0.004** | −0.005*** | −0.006*** | −0.004*** |
| | (0.005) | (0.002) | (0.002) | (0.002) | (0.002) | (0.002) | (0.002) | (0.002) |
| $export:s_{in}$ | 0.001 | 0.003*** | 0.003*** | 0.004*** | 0.005*** | 0.004*** | 0.005*** | 0.005*** |
| | (0.002) | (0.001) | (0.001) | (0.001) | (0.001) | (0.001) | (0.001) | (0.001) |
| $export:ess$ | −0.006 | −0.009* | −0.011** | −0.005 | −0.000 | 0.003 | 0.002 | 0.003 |
| | (0.013) | (0.005) | (0.004) | (0.004) | (0.003) | (0.003) | (0.003) | (0.003) |
| $Constant$ | −2.707*** | −2.695*** | −2.811*** | −2.838*** | −3.223*** | −2.962*** | −3.199*** | −3.217*** |
| | (0.176) | (0.253) | (0.210) | (0.180) | (0.132) | (0.129) | (0.134) | (0.126) |
| Observations | 9,000 | 9,253 | 10,308 | 16,474 | 20,604 | 21,579 | 24,909 | 25,829 |
| $R^2$ | 0.629 | 0.611 | 0.622 | 0.621 | 0.630 | 0.629 | 0.629 | 0.640 |
| Adjusted $R^2$ | 0.609 | 0.591 | 0.604 | 0.609 | 0.621 | 0.620 | 0.621 | 0.633 |
| Residual Std. Error | 0.318 (df = 8522) | 0.323 (df = 8783) | 0.318 (df = 9835) | 0.321 (df = 15963) | 0.318 (df = 20087) | 0.322 (df = 21060) | 0.328 (df = 24378) | 0.330 (df = 25299) |
| F Statistic | 30.325*** (df = 477; 8522) | 29.449*** (df = 469; 8783) | 34.290*** (df = 472; 9835) | 51.360*** (df = 510; 15963) | 66.329*** (df = 516; 20087) | 69.002*** (df = 518; 21060) | 77.851*** (df = 530; 24378) | 85.192*** (df = 529; 25299) |

**Table S18.** Cross sectional regressions for export. *** Significant at the 1% level. ** Significant at the 5% level. * Significant at the 10% level.



\end{document}